\documentclass[american,aps,pra,reprint,groupedaddress,twocolumn,showpacs]{revtex4-1} 
\usepackage[unicode=true,pdfusetitle, bookmarks=true,bookmarksnumbered=false,bookmarksopen=false, breaklinks=false,pdfborder={0 0 0},backref=false,colorlinks=false] {hyperref}
\hypersetup{ colorlinks,linkcolor=myurlcolor,citecolor=myurlcolor,urlcolor=myurlcolor}
\usepackage{braket,colortbl,cleveref,amsthm,amsmath,amssymb,txfonts}
\definecolor{myurlcolor}{rgb}{0,0,0.7}
\usepackage{graphics,graphicx}
\usepackage{color}
\usepackage{colortbl}
\usepackage{subfigure}
\usepackage{graphicx}
\usepackage[font=small,labelfont=bf,
   justification= centerlast,
   format=hang]{caption}
\usepackage{ragged2e}
\usepackage{amsmath}
\usepackage{graphicx}
\usepackage[utf8x]{inputenc}
\usepackage{color}
\usepackage{amsmath}
\usepackage{tikz}
\usetikzlibrary{shapes.geometric, arrows}
\usepackage{latexsym}
\usepackage{amssymb}
\usepackage{amsthm}
\usepackage{bm}
\usepackage{graphics,epstopdf}
\usepackage{color}\usepackage{amsmath}

\usepackage[usenames,dvipsnames,svgnames]{pstricks}
\usepackage{epsfig}
\usepackage{pst-grad} % For gradients
\usepackage{pst-plot} % For axes
\tikzstyle{startstop} = [rectangle, rounded corners, minimum width=3cm, minimum height=1cm,text centered, draw=black, fill=red!30]

\tikzstyle{env}=[circle,  ball color = green!20, minimum size= 80mm]
\tikzstyle{central}=[circle, ball color = red!100, minimum size=8mm]
\tikzstyle{bath}=[circle, ball color =blue!75, minimum size=4mm]

\theoremstyle{plain}

\def\bea{\begin{eqnarray}}
\def\eea{\end{eqnarray}}
\def\ba{\begin{array}}
\def\ea{\end{array}}

\def\beq{\begin{equation}}
\def\eeq{\end{equation}}

\begin{document}

\title{Quantum uncertainty relation based on the mean deviation }

\author{Gautam Sharma}
\email{gautamsharma@hri.res.in}
\affiliation{Quantum Information and Computation Group, Harish-Chandra Research Institute, Homi Bhabha National Institute, Allahabad 211019, India }

\author{Chiranjib Mukhopadhyay}
\email{chiranjibmukhopadhyay@hri.res.in}
\affiliation{Quantum Information and Computation Group, Harish-Chandra Research Institute, Homi Bhabha National Institute, Allahabad 211019, India }
 
\author{Sk Sazim}
\email{sksazim@hri.res.in}
\affiliation{Quantum Information and Computation Group, Harish-Chandra Research Institute, Homi Bhabha National Institute, Allahabad 211019, India }

\author{Arun Kumar Pati}
\affiliation{Quantum Information and Computation Group, Harish-Chandra Research Institute, Homi Bhabha National Institute, Allahabad 211019, India }
%\affiliation{Homi Bhabha National Institute, Training School Complex, Anushakti Nagar, Mumbai 400 085, India}

%  Arun Kumar Pati$^{1}$ \footnote{akpati@hri.res.in}}
% \affiliation{$^{1}$Harish-Chandra Research Institute, Chhatnag Road, Jhunsi, Allahabad - 211019, India \\ 
% and Homi Bhabha National Institute, Training School Complex, Anushakti Nagar, Mumbai 400 085, India}

%\author{Arun Kumar Pati$^{3}$ \footnote{...}}
%\affiliation{$^{3}$Harish-Chandra Research Institute, Chhatnag Road, Jhunsi, Allahabad - 211019, India \\}

\begin{abstract}
\noindent  Traditional forms of quantum uncertainty relations are invariably based on the standard deviation. This can be understood in the historical context of simultaneous development of quantum theory and mathematical statistics. Here, we present alternative forms of uncertainty relations, in both state dependent and state independent forms for a general set of deviation measures, with a special emphasis on the mean deviation. We illustrate the robustness of this formulation in situations where the standard deviation based uncertainty relation is inapplicable. We apply the mean deviation based uncertainty relation to detect EPR violation in a lossy scenario for a higher inefficiency  threshold than that allowed by the standard deviation based approach. We demonstrate that the mean deviation based uncertainty relation can perform equally well as the standard deviation based uncertainty relation as non-linear witness for  entanglement detection. 
%
%We also derive a quantum speed limit for unitary evolution based on the mean deviation based uncertainty relation - which although weaker than the existing bounds, suggest an intriguing bridge between the average energy based speed limits and variance based speed limits in literature. 

%\vspace{0.5cm}

%\textbf{ PACS numbers:} 

\end{abstract}

\maketitle

\section{Introduction}

%Quantum world has many puzzling attributes which demarcate it from the classical world. One such is 
%the `uncertainty principle'. After its inception by Heisenberg in 1927, there has been vast developments of 
%the theory which still persist till date \cite{CurrSc107.210}. The importance of such an attribute has widely been studied 
%and promalgated in many areas of science \cite{Folland1997}.
%
%Heisenberg's intuition has been formulated mathematically by Roberson in 1929 in the form of `uncertainty relation' which for 
%two observables reads as 
%\begin{equation}
% \Delta A \Delta B \geq \frac{1}{2}|\langle \psi|[A,B]|\psi\rangle|,
% \label{HRUP}
%\end{equation}
%where $\Delta \mathcal{O}=\sqrt{\langle\psi|(\mathcal{O}-\langle \psi|\mathcal{O}|\psi\rangle)^2|\psi\rangle}$ 
%is the standard deviation of the obsevable $\mathcal{O}$ in state $|\psi\rangle$. Roberson's relation tells us that 
%one cannot prepare a quantum ensemble in which two noncommuting observables can be measured with 
%infinite precision. However this relation does not capture all aspects of the `uncertainty principle' \cite{FBusc}.
%
%In Eq.\ref{HRUP}, the standard deviation has been considered as a measure of uncertainty of the observable 
%in a quantum state. However there exist other measures of dispersion in statistical theory, eg., `mean deviation' which 
%is obtained by considering the weighted sum of the absolute values of the deviations around the mean. 
One of the distinguishing features setting the quantum world apart from our daily experiences is the existence of fundamental uncertainty relations in the  former. Such uncertainty relations were first promulgated in mid 1920-s to mid 1930-s \citep{heisenberg, kennard, robertson}, coinciding with the era when modern mathematical works on statistics were beginning to proliferate \citep{fisherbook}.  In statistical analysis, at least until the advent of Shannon's theory of information, \emph{standard deviation (SD)} became the favored measure of dispersion due in no small part to the ease of manipulations and reliance on Gaussian models, inspired by the then recently developed central limit theorem \citep{Feller-book}. Thus, it is of little wonder that the uncertainty relations, dealing with the intrinsic spread associated to the probabilistic theory of quantum mechanics, were expressed in the mathematical form of standard deviations. Despite newer and ongoing developments on expressing uncertainty relations in entropic terms (See, for example, Ref.\citep{eur} for a rather detailed review on entropic uncertainty relations and their applications), text-book versions of the uncertainty relation still retain the standard deviation based expression \citep{sakurai-book}. In fact, interest in the standard deviation form of uncertainty relation has experienced a resurgence of late, with uncertainty relations beyond the Robertson-Schr{\"o}dinger inequality being proved \cite{macconepati,newurpaper} and generalized \cite{mult,newurpaper} for multiple observables. Subsequently, tighter and reverse uncertainty relations have also been proved using standard deviation based uncertainties \cite{shrobona}. However, standard deviation is not the only non-entropic measure of dispersion known \citep{bera_median}. The spread of probability distribution, quantified as average distance of an element of a probability distribution from the mean, can also be quantified, among other measures, by the \emph{mean deviation (MD)} \citep{giri, gorard} \footnote{Indeed, both SD and MD can be expressed as  distance measures in this view - SD as the Euclidean distance, MD as the $l_1$-distance.}. Just as the entropic formulation of quantum uncertainty relations has equipped us with novel insights and applications \citep{berta-nature,eurwehner}, it is natural to wonder whether any hidden gems lie within the hitherto less-explored uncertainty relations based on the pre-Shannon alternatives to the SD.
\\

In this work, we formulate uncertainty relations for quantum systems in terms of the mean deviation. Beginning with deriving the state dependent MD based uncertainty relation, we investigate the intelligent states in this framework and go on to derive a state independent uncertainty relation in terms of sum of MD based uncertainties for multiple observables. We illustrate that the formalism of MD based uncertainty relations is more robust than the usual variance based uncertainty relation with demonstration of examples of quantum states for which SD based uncertainty relations are inapplicable. As testament to the power of the new uncertainty relation, we show that the bound on efficiency of detection of EPR violation in a bipartite Werner-type state derived from our uncertainty relation is better than the bound obtainable through variance based uncertainty relations \citep{cavalcanti}. Moreover, we find that for this task, the mean deviation based uncertainty measure fares better than a large family of generalized uncertainty measures as well. We show that the mean deviation based uncertainty relation is an equally powerful tool for detection of entanglement in the bipartite scenario when compared to the variance based uncertainty relations \citep{hofmann, guehne}. 
%We further derive a speed limit for unitary quantum evolution from the mean deviation based uncertainty relation. While not as strong as the Mandelstam-Tamm speed limit obtainable through the variance based uncertainty relations \citep{deffnerreview}, our expression for the quantum speed limit is expressible in terms of the average energy, thus indicating a potential conceptual bridge between Margolus-Levitin type speed limits \citep{margolus, giovanetti} and uncertainty based speed limits \citep{mandelstam, anandan, pati, vaidman}  in literature. 

The paper is organized as follows. Sec.  \ref{II} presents the definition and proof of mean deviation based as well as generalized deviation measure based uncertainty relations. We specialize to the mean deviation based uncertainty relation for the analysis of intelligent states and derive a state independent mean deviation based uncretainty relation. We illustrate in Sec. \ref{III} the efficacy of our uncertainty relation in cases where the variance based uncertainty relations are inapplicable.  In Sec. \ref{IV},  we apply our uncertainty relations for detection of EPR steerability and entanglement. Finally, in Sec.\ref{V}, we conclude through indicating possible future developments and applications of our results.

\section{Mean deviation based uncertainty relations}
\label{II}

We first define the mean deviation based uncertainty for an observable  with respect to a quantum state.

\emph{Definition (MD based uncertainty)} - \textcolor{black}{For any physical observable $A=\sum_{a}a\ket{a}\bra{a}$, the mean deviation based uncertainty of an observable $A$ on the state $\ket{\Psi}$ is defined as} 
\bea
\Delta_M A &=  \sum_a|a-\langle A \rangle| |\bra{\Psi}\ket{a}|^2 \nonumber \\& =\bra{\Psi}{A^{\prime}}^2\ket{\Psi}.
\label{uncerA}
\eea

\noindent Note that $\Delta_M A$ is always non-negative and vanishes only when $|\Psi\rangle$ is an eigenstate of the observable $A$. Let us define a positive, Hermitian operator $A^{\prime} = \sum_a \sqrt{|a - \braket{a}|} \ket{a}\bra{a}$, and hence ${A^{\prime}}^2=\sum_a|a-\langle A \rangle|\ket{a}\bra{a}$. Similarly for the operator $B= \sum_b b\ket{b}\bra{b}$, we define $B^{\prime}$ and write down the uncertainty of $B$ \textcolor{black}{on} the state $\ket{\Psi}$ as 
\bea
\Delta_M B  &=  \sum_b|b-\langle B \rangle| |\bra{\Psi}\ket{b}|^2 \nonumber \\& =\bra{\Psi}{B^\prime}^2\ket{\Psi}.
\label{uncerB}
\eea

\noindent Let us now define two vectors $|\Psi_1 \rangle$ = $A' \ket{\Psi}$, and $\ket{\Psi_2} = B' \ket{\Psi}$, then we have $||\Psi_1||^2 = \langle \Psi | A'^2 | \Psi \rangle = \Delta_M A$ and $||\Psi_2||^2 = \langle \Psi | B'^2 | \Psi \rangle = \Delta_M B$.  Now, the product of $\Delta_M A$ and $\Delta_M B$ on the state $|\Psi\rangle $ respects the following inequality

\begin{eqnarray}
\Delta_M A \Delta_M B  \geq  \frac{1}{4}|\bra{\Psi}[A^{\prime},B^{\prime}]\ket{\Psi}|^2.
\label{mdur}
\end{eqnarray}
Here the  inequality follows from the Cauchy-Schwarz inequality for two unnormalized vectors $\ket{\Psi_1}$ and $\ket{\Psi_2}$. This is the Robertson form of  mean deviation uncertainty relation for products of uncertainties. It is also possible to cast the uncertainty relation in sum form instead of the product form above. For incompatible observers, the triviality of the SD based uncertainty relations was removed rather recently \cite{macconepati}.  Below we present a similar uncertainty relation in terms of the mean deviation.

\emph{Theorem (MD  based uncertainty relation for incompatible observables)} -  \textcolor{black}{For observables $A$ and $B$, system state $\ket{\Psi}$ and any state  $\ket{\Psi^\perp}$ orthogonal to the system state, the following uncertainty relation holds}  -\begin{equation} 
\Delta_M A + \Delta_M B  \geq \pm i\braket{[A^{\prime},B^{\prime}]} + |\bra{\Psi}A^{\prime}\pm iB^{\prime}\ket{\Psi^\perp}|^2.
\label{mdsur}
\end{equation} Here we choose the sign outside the commutator in such a way that the first term in the RHS remains positive.

\emph{Proof-} We write $\Delta_M A=||A^{\prime}\ket{\Psi}||$ and $\Delta_M B=||iB^{\prime}\ket{\Psi}||$ to obtain
\beq
||(A^{\prime}\mp iB^{\prime})\Psi||^2=\Delta_M A +\Delta_M B \mp i\braket{[A^{\prime},B^{\prime}]}. \nonumber
\eeq
	
\noindent Now the LHS of this expression can be lower bounded using the Cauchy-Schwarz inequality as  $||(A^{\prime}\mp iB^{\prime})\Psi||^2 \geq |\bra{\Psi} A^{\prime}\pm iB^{\prime}\ket{\Psi^{\perp}}|^2$, for every $\ket{\Psi^{\perp}}$ orthogonal to $\ket{\Psi}$. This completes the proof. \qed

Similar to the Robertson-Schr{\"o}dinger uncertainty relation, the MD based uncertainty relation given in \eqref{mdur} can be trivial when  $\ket{\Psi}$ is an eigenstate of either $A$ or $B$.  However, the lower bound in \eqref{mdsur} is non-trivial for every $\ket{\Psi^{\perp}}$, barring the case when $\ket{\Psi}$ is a common eigenstate of both $A^{\prime}$ and $B^{\prime}$.  
\subsubsection{Case for generalized deviation measure}

The obvious generalization of the mean deviation based uncertainty relation derived above would be to consider the situation for arbitrary exponent $\alpha$, which would subsume the mean deviation based uncertainty measure and the usual variance based uncertainty measure as special cases, viz., when $\alpha = 1$ or $\alpha = 2$ respectively. More concretely, similar to Eq.\eqref{uncerA}, we seek to define generalized deviations as
\begin{align}
\Delta_M^{\alpha}A= &\sum_{a} |\bra{\Psi}\ket{a}|^2 |a-\braket{A}|^{\alpha} \nonumber \\ & = \bra{\Psi}{A_{\alpha}^{\prime}}^2\ket{\Psi}.
\label{alphauncerA}
\end{align}
\noindent where $\lbrace a\rbrace$ is the set of eigenvalues of the observable $A$ and $\lbrace\ket{a} \rbrace$ are the corresponding eigenvectors. As done earlier let us define a positive semi-definite operator $A_{\alpha}^{\prime} = \sum_a \sqrt{|a - \braket{a}|^{\alpha}} \ket{a}\bra{a}$. Hence, ${A_{\alpha}^{\prime}}^2=\sum_a|a-\langle A \rangle|^{\alpha}\ket{a}\bra{a}$. For the operator $B= \sum_b b\ket{b}\bra{b}$, we similarly define $B_{\alpha}^{\prime}$ and write down the uncertainty of $B$ \textcolor{black}{on} the state $\ket{\Psi}$ as 
\bea
\Delta_M B  &=  \sum_b|b-\langle B \rangle|^{\alpha} |\bra{\Psi}\ket{b}|^2 \nonumber \\& =\bra{\Psi}{B_{\alpha}^\prime}^2\ket{\Psi}.
\label{alphauncerB}
\eea

The resulting product and sum uncertainty relations are now expressed as 
\begin{eqnarray}
\Delta_M^{\alpha} A \Delta_M^{\alpha} B  \geq  \frac{1}{4}|\bra{\Psi}[A^{\prime}_{\alpha},B^{\prime}_{\alpha}]\ket{\Psi}|^2,
%\label{mdur}
\end{eqnarray}
and 
\begin{equation} 
\Delta_M^{\alpha} A + \Delta_M^{\alpha} B  \geq \pm i\braket{[A^{\prime}_{\alpha},B^{\prime}_{\alpha}]} + |\bra{\Psi}A^{\prime}_{\alpha}\pm iB^{\prime}_{\alpha}\ket{\Psi^\perp}|^2,
%\label{mdsur}
\end{equation}
respectively.

\subsection{Intelligent states}

It is natural  to wonder which quantum states are the most `classical' in the sense of incurring the least amount of uncertainty for incompatible observables. A canonical example is that of the coherent states for a quantum harmonic oscillator \citep{sakurai-book}. These states have been given the moniker of `intelligent' states in the literature and studied for the SD based uncertainty relations \cite{pati}.  It is well known that a Gaussian wavefunction saturates the uncertainty bound of standard deviation based uncertainty relation. Specifically, for the position and momentum operators the lower bound is given by 
\begin{align*}
\Delta X \Delta P=\frac{\hbar}{2}.
\end{align*} 
Here the SD for the position observable is defined as $\Delta X = \sqrt{\bra{\Psi} X^2 \ket{\Psi} - \left(\bra{\Psi} X \ket{\Psi}\right)^{2}}$ and SD of momentum similarly.  However, if we move away from the SD based approach, the situation is less clear. For median based uncertainty relations, it was numerically shown  \citep{bera_median} that the wave function corresponding to  the Cauchy probability distribution is, in fact, more `intelligent' than the Gaussian wave function, which is not reflected in the SD based uncertainty relations, owing to the fact that the SD (or indeed even the mean) does not exist in general for the Cauchy type probability distribution. One can easily calculate the product of mean deviation uncertainties in position and momentum for the Gaussian wavefunction and the product is given by ($\hbar = 1$)
\begin{align*}
\Delta_M X \Delta_M P=\frac{1}{\pi} .  \hspace{0.4in} 
\label{gaussianintel}
\end{align*}
The only difference in this case as compared to the SD  is the factor $\pi$ in the denominator. 

 In the quest for finding intelligent states in the MD case, let us now digress a bit. The differential entropy was introduced by Shannon himself in a bid to generalize the Shannon entropy for continuous settings. It is defined as follows-

\emph{Definition (differential entropy)} - \textcolor{black}{If $X$ is a random variable with a probability density function $p$ whose support is the set $\mathbb{X}$, then the  differential entropy $H(X)$ is defined as \beq H(X) = - \int_{\mathbb{X}} p(x) \ln p(x) \ dx .\eeq }

It can be shown that the probability distribution that maximizes the differential entropy given a fixed SD is a Gaussian. We now prove the analog of this result for the MD case.  

\emph{Theorem (Probability density function which maximizes differential entropy for a fixed value of mean deviation)} - \textcolor{black}{The Laplace distribution maximizes the differential entropy if the mean deviation is fixed (say $\mu$) and the mean is set to zero. }
\begin{proof}
We have to maximize $H(x)$ subject to the following constraints
\begin{enumerate}
\item  $\mu=\int p(x)|x| dx,$
\item  $ \int p(x) dx=1.$
\end{enumerate}
Now introducing the Lagrange multipliers $\lambda$ and $\gamma$, the functional derivative of the following quantity $\int [-p(x)\ln p(x)+\lambda |x|p(x) +\gamma p(x)]dx $ must vanish for maximization, which immediately leads to the result
\begin{equation}
 -1-\ln p(x)+\lambda |x|+\gamma=0.
\end{equation}
Utilizing the constraints to eliminate the Lagrange multipliers, we end up with the following Laplace probability density function, i.e.,
\begin{equation}
 p(x)=\frac{1}{2 \mu}\exp\bigg(-\frac{|x|}{\mu}\bigg). \nonumber 
\end{equation} 
\end{proof}
\textcolor{black}{This constrained optimization procedure  bears a strong resemblance to the way one singles out the Gibbs distribution by fixing the average energy and maximizing the von Neumann entropy. Here we begin by fixing the mean deviation, which is a measure of dispersion, unlike the average energy.} However, the above result immediately spawns the question - are the wave functions giving rise to the Laplace probability distribution as `intelligent' as the Gaussian wave function as far as the MD based uncertainty relation is concerned ? We answer this question in the affirmative through the following proposition. 

\emph{Proposition (States as intelligent as Gaussian states in the context of the MD uncertainty relation)} - \textcolor{black}{States with wave function generating the Laplace probability distribution are as intelligent as Gaussian states in the context of the MD uncertainty relation. }

\emph{Proof-} We assume the following position space wave function ($\hbar =1$)
\begin{equation}\psi(x)= \frac{1}{\sqrt{2\mu}}\exp\bigg(-\frac{|x|}{2 \mu}\bigg).
\end{equation}\noindent The mean deviation corresponding to the above wavefunction is given by $\Delta_M X= \mu.$ The momentum space wave function, i.e., the Fourier transform of the position space wave function is a Cauchy distribution \beq \tilde{\psi}(p)=\frac{2\sqrt{\mu}}{\sqrt{\pi}(1+4\mu^2p^2)}. \eeq \noindent The mean deviation for momentum is, therefore, given by $ \Delta_M P= \frac{1}{\pi\mu}.$ Hence, the product of MD based uncertainties in position and momentum reads as 
\begin{equation}
 \Delta_M X \Delta_M P= \frac{1}{\pi}.
\end{equation}
This is exactly the expression for Gaussian wave function given earlier, thus completing our proof. \qed
\\
We note in passing that the above wave function arises naturally as a solution of the Schr{\"o}dinger equation for the one-dimensional Dirac delta potential \citep{Ballentine-book}. However, the corresponding state is not as `intelligent' as the Gaussian state for the product of SD based uncertainties as it satisfies  $\Delta X \Delta P = \frac{1}{\sqrt{2}}$, whereas in the Gaussian case, $\Delta X \Delta P = \frac{1}{2}$.  One is tempted to ask the question - are these the most intelligent states in the MD scenario ? The answer to this question is quite tricky, as has been pointed out in an online forum \citep{stack} by Frederic Grosshans. He showed, if one uses the entropic uncertainty relation for position and momentum in conjunction with the fact that the differential entropy is related to the mean deviation, one can get a bound on the product of the MD uncertainties, which is somewhat lower than $\frac{1}{\pi}$.  This bound is saturated in the case that both the position space and momentum space wave functions generate Laplace probability distributions - which is not possible since the Fourier transform of a Laplace distribution is not another Laplace distribution. Thus, we leave the problem of finding more intelligent states than the ones discussed in this section for future work.

\subsection{State independent MD based uncertainty relation }

One of the nice features of entropic uncertainty relations is the fact that they are state independent and consequently bring out the incompatibility of pairs of observables without having to worry about states for which the uncertainty relation becomes trivial. It is thus a natural question to ask whether we can have state independent uncertainty relations for other measures of uncertainty. For SD, this was addressed by Huang \citep{huang}. In this subsection, we provide a state independent MD based uncertainty relation for the sum of an arbitrary number of observables.

\emph{Setting-} Suppose we construct $m$ number of bases $\lbrace\mathcal{B}_i \rbrace_{i = 1 ... m}$ for an $n$-dimensional Hilbert space. Now
let $|a_i^j\rangle $ be the $j$-th basis element for the $i$-th basis. We now assume $m$ Hermitian operators of the form $\mathcal{O}_i=\sum_ja_i^j|a_i^j\rangle\langle a_i^j| $ . We consider the probabilities corresponding to the measurement outcome of observables as $p_i^j=|\langle a_i^j|\Psi\rangle|^2$, where $|\Psi\rangle$ is the corresponding state.  The aim is to provide a state independent lower bound for the sum of the MD based uncertainties of these observables. To this end, we formulate the following uncertainty relation

\emph{Theorem (state independent MD based uncertainty relation) -- } The following MD based sum uncertainty relation holds for multiple observables $\lbrace \mathcal{O}_i \rbrace$ and any $\alpha \in \mathbb{R}$  \beq 
\alpha \sum_i\Delta_M(\mathcal{O}_i)\geq  
 C -\sum_i\ln\max_{\min_k a_i^k\leq \beta_i\leq\max_k a_i^k}\sum_je^{-\alpha |a_i^j
-\beta_i |} .
 \eeq Here, $C$ in the lower bound is the logarithm of the maximum overlap \citep{maassenuffink} between operator spectra, and consequently is state independent.  To prove this result, we first consider the following lemma. 

\emph{Lemma} - \textcolor{black}{For $\alpha\in \mathbb{R}$,
\begin{equation}
 \alpha \Delta_M(\mathcal{O}_i)\geq H(\mathcal{O}_i)
 -\ln\sum_je^{-\alpha |a_i^j-\langle\mathcal{O}_i \rangle|}.
 \label{lemm1}
\end{equation}
where $H(\mathcal{O}_i)=-\sum_j p_i^j\log_2 p_i^j$ is the Shannon entropy 
of the observable $\mathcal{O}_i$and $\langle\mathcal{O}_i \rangle$ is its mean}. \\

\emph{Proof}- Using the inequality $e^x\geq 1+x$ in conjunction with $x=-\alpha|a_i^k-\langle\mathcal{O}_i \rangle|- 
\ln p_i^k \sum_je^{-\alpha |a_i^j-\langle\mathcal{O}_i \rangle|}$, we find that
\begin{eqnarray}
 1 &=& \sum_k p_i^k \frac{e^{-\alpha |a_i^k-\langle\mathcal{O}_i \rangle|}}
 {p_i^k \sum_je^{-\alpha |a_i^j-\langle\mathcal{O}_i \rangle|}},\nonumber\\
 & \geq & \sum_k p_i^k (1-\alpha|a_i^k-\langle\mathcal{O}_i \rangle|- 
\ln p_i^k \sum_je^{-\alpha |a_i^j-\langle\mathcal{O}_i \rangle|}),\nonumber\\
&=& \sum_k p_i^k -\alpha\sum_k p_i^k|a_i^k-\langle\mathcal{O}_i \rangle| -\sum_k p_i^k\ln p_i^k 
\nonumber \\ && -\sum_k p_i^k\ln \sum_je^{-\alpha |a_i^j-\langle\mathcal{O}_i \rangle|},\nonumber\\
&=& 1 -\alpha \Delta_M(\mathcal{O}_i)+ H(\mathcal{O}_i) 
-\ln\sum_je^{-\alpha |a_i^j-\langle\mathcal{O}_i \rangle|}.
\end{eqnarray} \noindent This completes the proof of the lemma. \qed

\noindent Now summing over $i$ in \eqref{lemm1}, we get the state dependent MD based sum uncertainty relation. 
\begin{eqnarray}
 \alpha \sum_i\Delta_M(\mathcal{O}_i)&\geq & \sum_iH(\mathcal{O}_i)
 -\sum_i\ln\sum_je^{-\alpha |a_i^j-\langle\mathcal{O}_i \rangle|},\nonumber\\
 &=& C -\sum_i\ln\sum_je^{-\alpha |a_i^j-\langle\mathcal{O}_i \rangle|}. 
\end{eqnarray}
Now, noticing that $\min_k a_i^k\leq \langle\mathcal{O}_i \rangle\leq\max_k a_i^k$, we find
\begin{eqnarray}
 \alpha \sum_i\Delta_M(\mathcal{O}_i)&\geq & 
 C -\sum_i\ln\max_{\min_k a_i^k\leq \beta_i\leq\max_k a_i^k}\sum_je^{-\alpha |a_i^j
-\beta_i |}.\; \; \;
\label{state_ind}
\end{eqnarray}
% where $Q=\sum_i\ln\max_{\min_k a_i^k\leq \beta_i\leq\max_k a_i^k}\sum_je^{-\alpha |a_i^j
% -\beta_i |}$. 
\noindent This accomplishes the goal of finding a MD based state independent uncertainty relation for multiple observables.
\vspace{0.25in}
\section{New Uncertainty Relation: Examples }
\label{III}
The constraints placed on quantum wave functions are less severe than restricting the possible solutions of the Schr{\"o}dinger equation only to functions for which the standard deviation does not blow up at any point. This gives rise to perfectly licit solutions of the Schr{\"o}dinger equation, for which the standard deviation based uncertainty relations are inapplicable. One way of dealing with this problem is to resort to the \emph{semi interquartile range (SIQR)} as a measure of spread \citep{bera_median}. However, analytical expressions for SIQR of arbitrary distributions are notoriously hard to calculate. As we argue below, the MD based uncertainty relation \eqref{mdur} derived above is algebraically more tractable and as such, an excellent candidate to fill this lacuna. These relations only demand that the mean be well-defined, which is a less stringent condition than requiring an well-behaved standard deviation. In this section, we illustrate two different scenarios where, in some regimes, the SD based uncertainty relations are inapplicable, but the new uncertainty relations hold true. We illustrate this using two examples, one being the F-distribution, the other being the Pareto distribution.

\subsection{F-distribution}
Let us now consider the probability distribution function known as F-distribution, whose expression is given by
\begin{align}\label{fpdf}
f(x;d_1,d_2)=\frac{1}{\beta(\frac{d_1}{2},\frac{d_2}{2})}\left(\frac{d_1}{d_2}\right)^{\frac{d_1}{2}}x^{\frac{d_1}{2}-1}\left(1+\frac{d_1}{d_2}x\right)^{-\frac{d_1+d_2}{2}}.
\end{align}

\noindent where $x\geq0$,  $\beta(a,b)$ being the two-parameter beta function family and the parameters $d_1$ and $d_2$ being positive integers. The mean, given by $\frac{d_2}{d_2-2}$, exists for all $d_2>2$. The standard deviation $\sigma=\sqrt{\frac{2{d_2}^2(d_1+d_2-2)}{d_1(d_2-2)^2(d_2-4)}}$, exists however only for $d_2 >4$. Thus, we note that the standard deviation for this distribution does not exist for $d_2= \lbrace 3,4 \rbrace$ even though the mean exists. 

It can be shown that this distribution arises as a solution to the Schr{\"o}dinger equation for the following form of potential ($V_0$ being a constant energy shift parameter).

\begin{widetext}
\begin{equation}
V(x)=V_0-\frac{\hbar^2}{32m}\left[(2d_1-2)(2d_1-6)x^{-2}-2\frac{d_1}{d_2}(2d_1-2)(d_1+d_2)x^{-1}\left(1+\frac{d_1}{d_2}x\right)^{-1}+\frac{d_1}{d_2}^2(d_1+d_2)(d_1+d_2+4)\left(1+\frac{d_1}{d_2}x\right)^{-2}\right].
\end{equation}
\end{widetext}

\begin{figure}
\subfigure[ Physical potential $V(x)$ as a function of position $x$ corresponding to the F-distribution probability density. Here we have set $V_0 = 0$ and $\hbar = m = 1$ and $d_1 =1$.]{
    \includegraphics[width=0.22\textwidth, keepaspectratio]{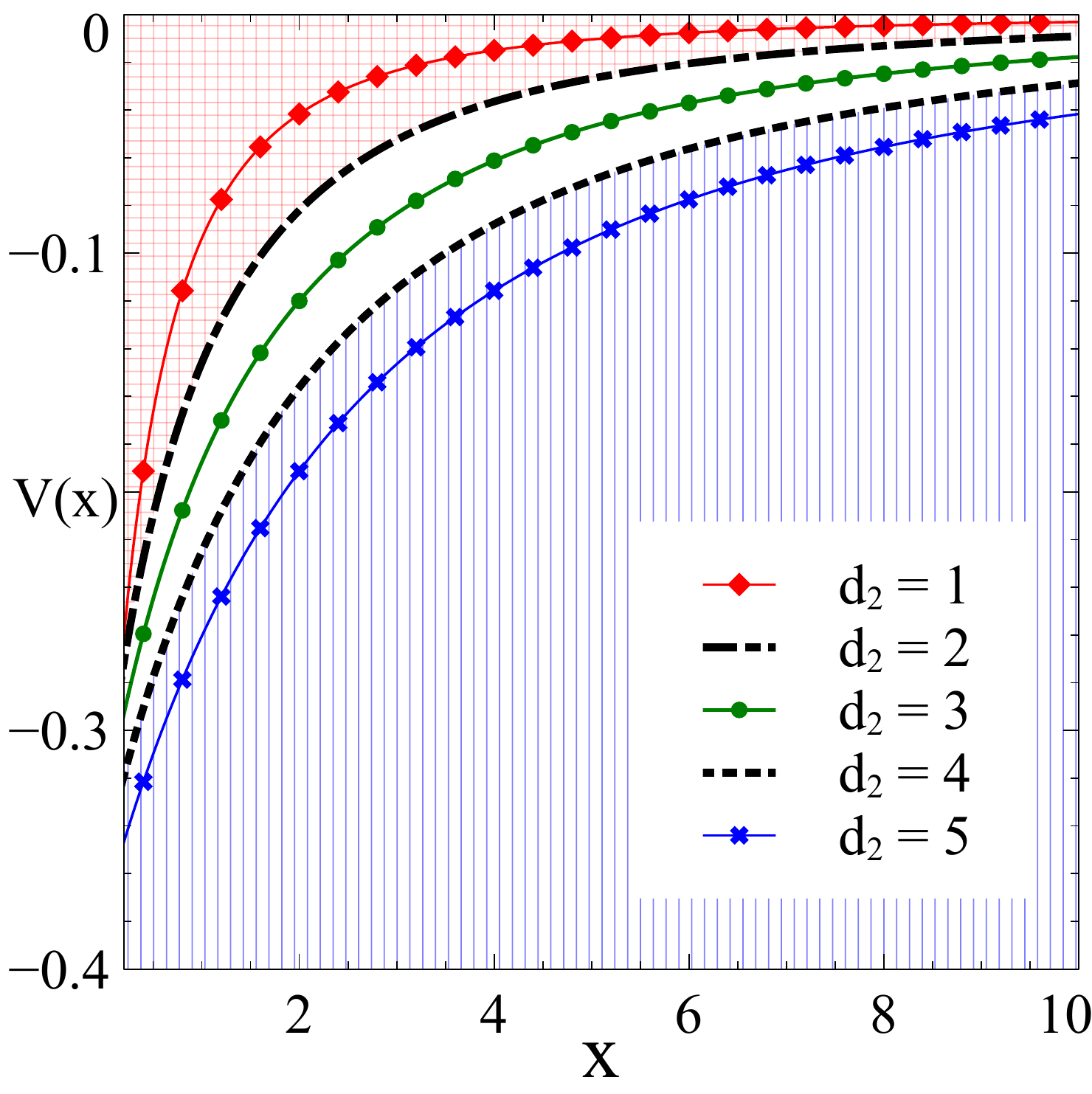}
    \label{f_potential}
}
\subfigure[ Probability distribution function for the F-distribution for various p values of $d_2$ ($d_1 =1$)]{
    \includegraphics[width=0.22\textwidth, keepaspectratio]{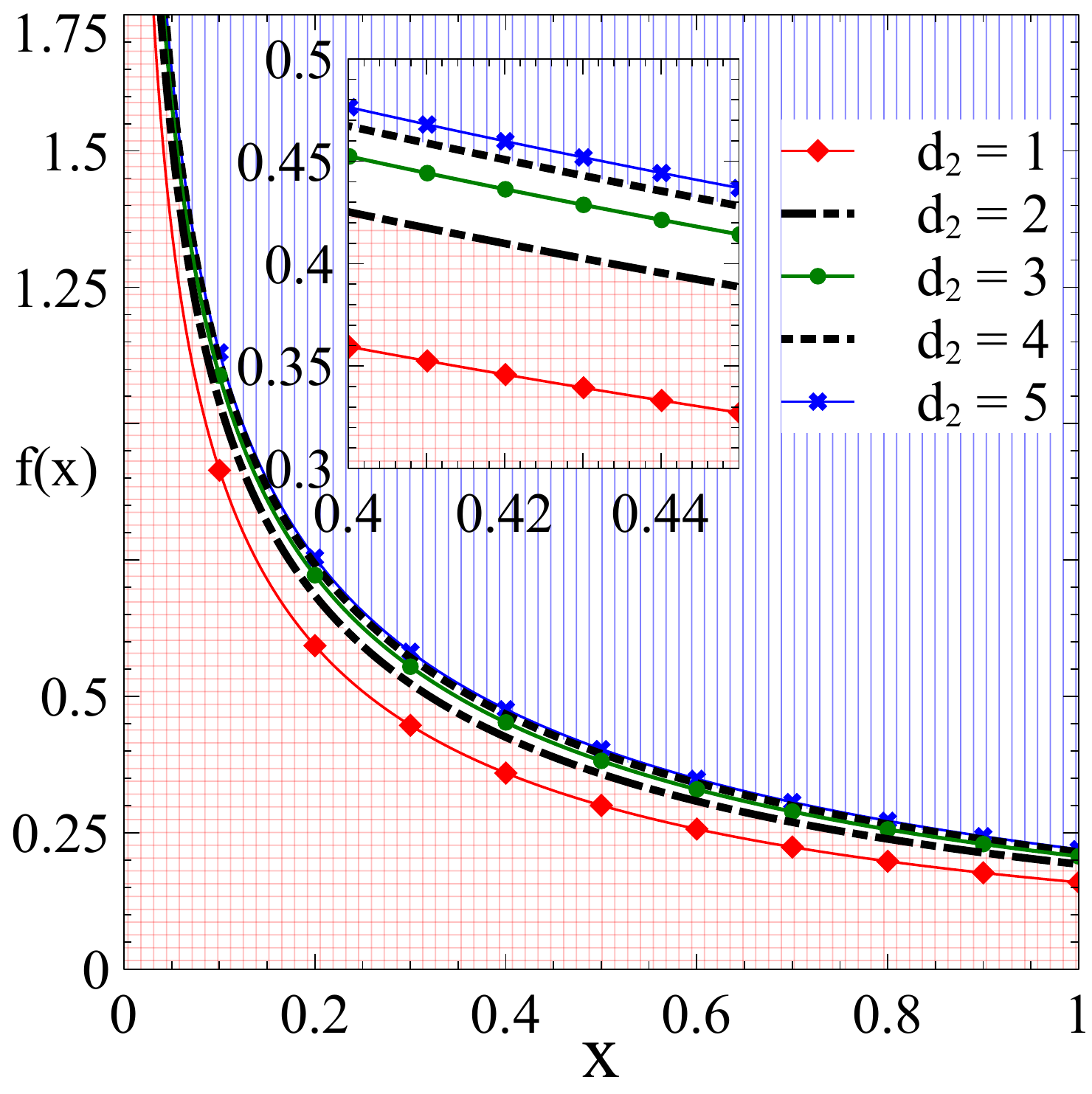}       
    \label{f_prob}
}

\caption{(Color online) Graphical depiction of the efficacy of MD based uncertainty relations. The blue striped zone, with an example depicted by the blue line corresponding to $d_2 = 5$, is where both SD based uncertainty relations and MD based uncertainty relations are applicable. The white zone, with an example depicted by the green line corresponding to $d_2 =3$, is where the MD based uncertainty relations apply but SD based ones do not. The red squared zone, with an example furnished by the red line corresponding to $d_2 =1$, is where both the MD and SD  based uncertainty relations fail to apply. The lines corresponding to $d_2 = 2$ (dot-dashed) and $d_2 = 4$ (dashed) set the boundaries beween these zones. We set $d_1 =1$ throughout.}
\label{f_distribution}
\end{figure}

It is clear that in the regime $d_2 \in (2,4]$, the SD-based uncertainty relation is meaningless. However, the mean deviation for F-distribution is perfectly defined in that regime(see Fig.\ref{f_distribution}) and is, in general, for $d_2 > 2$, given by
\begin{widetext}
\begin{align*}
\Delta_MX&=\frac{1}{B(\frac{d_1}{2},\frac{d_2}{2})^{\frac{1}{2}}}\left(\frac{d_1}{d_2}\right)^{\frac{d_1}{4}}[2\left(\frac{d_2}{d_2-2}\right)^{\frac{2+d_1}{2}}\left(\frac{{}_2F_1(\frac{d_1}{2},\frac{d_1+d_2}{2},\frac{d_1+2}{2},-\frac{d_1}{d_2-2})}{d_1}-\frac{{}_2F_1(\frac{d_1+2}{2},\frac{d_1+d_2}{2},\frac{d_1+4}{2},-\frac{d_1}{d_2-2})}{d_1+2}\right) \\& +\frac{2\left(\frac{d_2-2}{d_1}\right)^{\frac{d_2}{2}}\left(\frac{d_1}{d_2}\right)^{\frac{-d_1}{2}}\left((d_2){{}_2F_1(\frac{d_2-2}{2},\frac{d_1+d_2}{2},\frac{d_2}{2},\frac{2-d_2}{d_1})}-(d_2-2){{}_2F_1(\frac{d_2}{2},\frac{d_1+d_2}{2},\frac{d_1+2}{2},\frac{2-d_2}{d_1})}\right)}{(d_2-2)^2}].
\end{align*}
\end{widetext}

\noindent where ${}_2F_1(a,b,c,z)$ is the hyper-geometric function. \textcolor{black}{We acknowledge that this is not a potential that one comes across very often in literature. However, this is a possible physical potential, and may turn out to be relevant for future works.}
\subsection{Pareto distribution}

As another example of a physical situation where the mean deviation based uncertainty relation is meaningful in contradistinction with SD based uncertainty relations, let us assume a solution of the Schr{\"o}dinger equation of the form 

\begin{equation}\label{ppdf1}
\psi (x) =\left\{
\begin{array}{@{}ll@{}}
f(x) , & \text{if}\ x \geq \lambda  \\
\phi (x), & \text{otherwise}
\end{array}\right.
\end{equation}

\noindent where $f(x)$ arises from the Pareto distribution and is defined as  
\begin{align}\label{ppdf2}
f(x) = \sqrt{p} \sqrt{\frac{\alpha \lambda^{\alpha}}{x^{\alpha + 1}}}. 
\end{align}
where $p \in (0,1)$ and $\lambda \geq 0$.
In order to ensure the continuity of the wave function, the constraints on $\phi(x)$ are given by
\begin{itemize}
	\item i) $\phi (\lambda) = \sqrt{\frac{\alpha}{2 \lambda}}$, \\
	\item ii) $ \phi'(\lambda) = - \sqrt{\frac{\alpha}{2 \lambda}} \frac{\alpha + 1 }{2 \lambda} $,\\
	\item iii) $\int_{ - \infty}^{\lambda} |\phi (x) |^{2} dx = 1- p $.
\end{itemize}

Obviously, one can construct families of functions satisfying these properties. For each such function, the corresponding physical potential can be found. Now, it is easy to see that  the fluctuation in position for this wave function $\psi(x)$ blows up, yet the mean is well defined in the regime $\alpha \in (1,2] $ - thus the standard deviation based uncertainty relation becomes meaningless in this regime, yet the mean deviation based uncertainty relations are perfectly meaningful even in this scenario. Mean position $\beta = \langle X \rangle$ can now be easily seen to be finite for legitimate wave functions $\phi (x)$. Therefore, $\Delta_{M} X  = \int_{-\infty}^{\infty} |\psi(x)|^{2} |x - \langle X\rangle | dx = \int_{ - \infty}^{\lambda} |\phi (x)|^{2} |x - \beta|dx  + \int_{\lambda}^{\infty} |f(x)|^{2} |x - \beta| dx $. Now, both the terms are finite, therefore the mean deviation is finite in this case.

\section{Some applications of the mean-deviation based uncertainty relations}
\label{IV}

Apart from being one of the cornerstones of quantum theory, uncertainty relations can be applied to provide new insights on future quantum technologies. Uncertainty relations have successfully been utilized, among other applications, as (non-linear) entanglement witnesses \citep{hofmann,guehne}, in determining the speed limit of evolution of quantum states \citep{mandelstam, anandan, vaidman}, in determining the purity of states \citep{mal}, detecting EPR steering \citep{reid} and in determining the degree of non-locality of proposed physical theories through retrieval games \citep{wehner}. Till now, most of the tasks mentioned above have been performed using either the SD-based form or the entropic form of the uncertainty relations. Thus, it is natural to wonder how our mean deviation based uncertainty relations fare over standard deviation based uncertainty relations in these applications. In the present work, we seek to provide an answer in two such situations. First, we consider the problem of detecting EPR-steering. Finally, we analyze the utliity of mean deviation based uncertainty relations in entanglement detection. 

 \subsection{Detection of EPR violation}

%In Ref.\cite{cavalcanti}, the criteria to detect EPR paradox was given by using uncertainty relations based on variances. We prove that one can also give a similar criteria to detect EPR paradox using mean deviation based uncertainty relations. Also, it was shown in Ref.\cite{Cavalcanti2009} that by using variance based sum uncertainty relations that one can achieve EPR paradox for spin-entangled particles for an efficiency of $58\%$. However, by using the mean deviation based sum uncertainty relations we show that EPR paradox can be achieved for spin-entangled particles for an efficiency of $33\%$. \\

 \begin{figure}[htb]
 \includegraphics[scale=0.17]{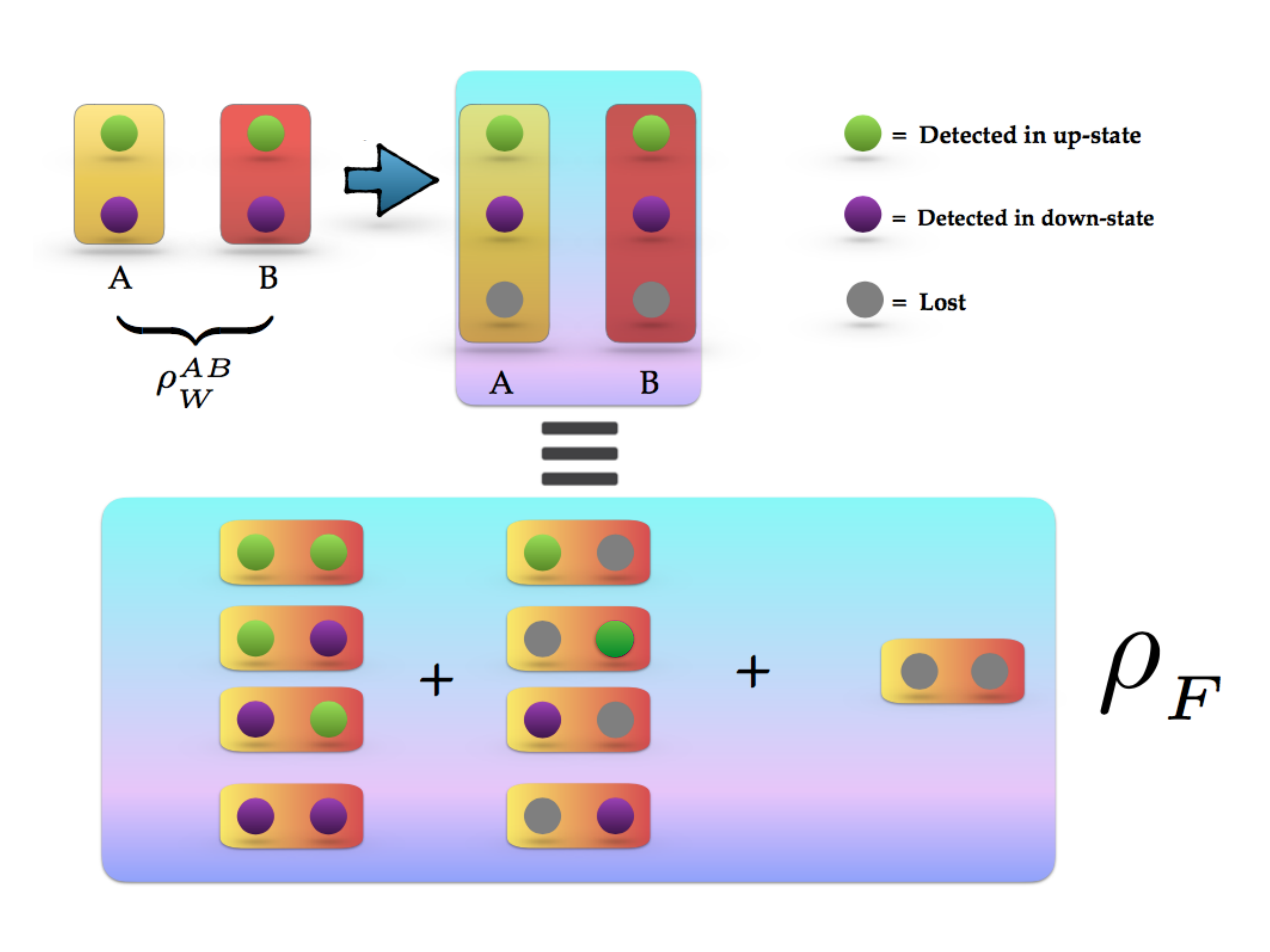}
 \caption{(Color online) A schematic diagram for the lossy detector scenario. The two-qubit initial Werner state is written in terms of a $9\times 9$ density matrix $\rho_F$ taking into account particle loss before measurement.}
 \label{schematic}
 \end{figure}

One of the oldest philosophical objections to the quantum theory is the EPR argument \citep {epr, bell} of local realism. Typically, constraining the theory to satisfy local realism at one or more subsystems results in certain inequalities \citep{bell, wiseman}, the violation of which for some quantum state implies the untenability of the EPR assumption. Here we consider the MD based local uncertainty relations for a subsystem $A$ of a bipartite state $\ket{\Psi}_{AB}$. For a set of observables $\lbrace \mathcal{O}_i \rbrace$, we define inferred mean deviations as $\Delta_{M_{\text{inf}}}\mathcal{O}_i=\sum_{\mathcal{O}^B}P(\mathcal{O}^B)\Delta_{M}(\mathcal{O}_i|\mathcal{O}^B)$  and inferred mean $|\braket{C}_{\text{inf}}|=\sum_{\mathcal{O}^B}P(\mathcal{O}^B)|\braket{C|\mathcal{O}^B}|$ and apply them for product uncertainty relations of the form $\Delta_M\mathcal{O}_1 \Delta_M\mathcal{O}_2\geq \frac{1}{4}|\bra{\Psi_A}[\mathcal{O}_1^{\prime},\mathcal{O}_2^{\prime}]\ket{\Psi_A}|^2=|\braket{C}|^2$ and corresponding sum uncertainty relations in the Robertson-like form. 

 \textit{Theorem (EPR violation)}-  \textcolor{black}{If we replace the mean deviations by the inferred mean deviations $\Delta_{M_{\text{inf}}}\mathcal{O}_i$, and the mean $|\braket{C}|$ by the inferred mean  $|\braket{C}_{\text{inf}}|$ in an uncertainty relation of the form $\Delta_M\mathcal{O}_1 \Delta_M\mathcal{O}_2\geq \frac{1}{4}|\bra{\Psi}[
 \mathcal{O}_1^{\prime},\mathcal{O}_2^{\prime}]\ket{\Psi}|^2=|\braket{C}|^2 $ - the violation of the resulting inequality is a manifestation of the EPR paradox.}
 
\textit{ Proof -} To demonstrate the EPR phenomena we assume using the local realism argument that there exists an element of reality $\lambda_{\mathcal{O}_i}$ which probabilistically predetermines the result for the measurement of the observable $\mathcal{O}_i$  performed at A. For two different elements of reality we can have a joint probability distribution of the form $P(\lambda_1,\lambda_2)= P(\lambda)$.  Now for the product of two mean deviations 
\begin{align*}
\Delta_{M_{\text{inf}}}\mathcal{O}_1 \Delta_{M_{\text{inf}}}\mathcal{O}_2 &= \sum_{\mathcal{O}_1^B}P(\mathcal{O}_1^B)\Delta_{M}(\mathcal{O}_1|\mathcal{O}_1^B) \sum_{\mathcal{O}_2^B}P(\mathcal{O}_2^B)\Delta_{M}(\mathcal{O}_2|\mathcal{O}_2^B) \\
&\geq \sum_{\lambda}P(\lambda)\Delta_{M}(\mathcal{O}_1|\lambda)\Delta_{M}(\mathcal{O}_2|\lambda) \\
&\geq \sum_{\lambda}P(\lambda)|\braket{C|\lambda}|^2\\
&\geq |\braket{C}_{\text{inf}}|^2.
\end{align*}
 
The EPR inequality for sum of mean deviations can be proved in a similar way as 
\begin{align}
\label{sumur}
\Delta_{M_{inf}}\mathcal{O}_1 + \Delta_{M_{inf}}\mathcal{O}_2  &= \sum_{i}\sum_{\mathcal{O}_i^B}P(\mathcal{O}_i^B)\Delta_{M}(\mathcal{O}_i|\mathcal{O}_i^B) \nonumber  \\
& =\sum_{\lambda}[\Delta_{M}(\mathcal{O}_1|\lambda) +\Delta_{M}(\mathcal{O}_2|\lambda)] \geq 2 |\braket{\textcolor{black}{D}}_{\text{inf}}| .
 \end{align} Here $\braket{D}$ corresponds to the first term in the rhs of the sum uncertainty relation \eqref{mdsur}. This completes the proof of the EPR inequality in terms of (inferred) mean deviation.  \qed
 
 In the next part of this subsection, we concentrate on applying the above result in an experimental setup.
 
\subsubsection{Detection of EPR violation in lossy scenario through MD uncertainty relation} 
Armed with the inequality \eqref{sumur} above, we now consider an experimental scenario for observing EPR violation upon measurement on one of the parties in a two-qubit setting.  Specifically, we work with the set-up proposed in Ref. \citep{Cavalcanti2009}, which we outline below for the sake of completeness. See Fig. \ref{schematic} for an illustration.

\noindent \emph{Scenario-}  Consider two spatially separated particles at locations A and B respectively, each of which can either be in spin-up (+1) or in spin-down (-1) configuration. Now assume, they share a singlet state $\ket{\Psi}_{AB}=\frac{1}{\sqrt{2}}(\ket{1}_A\ket{-1}_B-\ket{-1}_A\ket{1}_B)$ along with a white noise. Their shared state is thus described in general by the two qubit Werner family of states $\rho_W^{AB}=p\ket{\Psi}_{AB}\bra{\Psi}+\frac{1-p}{4} \mathbb{I}_4$.  Now, let us assume that a detector observes the spin of the each of the particles. However, this detector is inefficient in the sense that sometimes it may fail to conclusively detect a particle in either spin-up or spin-down configuration due to loss of that particle before measurement. This is parametrized by introducing an overall detection efficiency $\eta$. The detection space for each of the spins  is now that of a qutrit with possible outcomes being \begin{itemize}
\item{spin-up (+1),}
\item{spin-down (-1),}
\item{lost particle (0).}
\end{itemize}
\begin{figure}
 \includegraphics[scale=0.4]{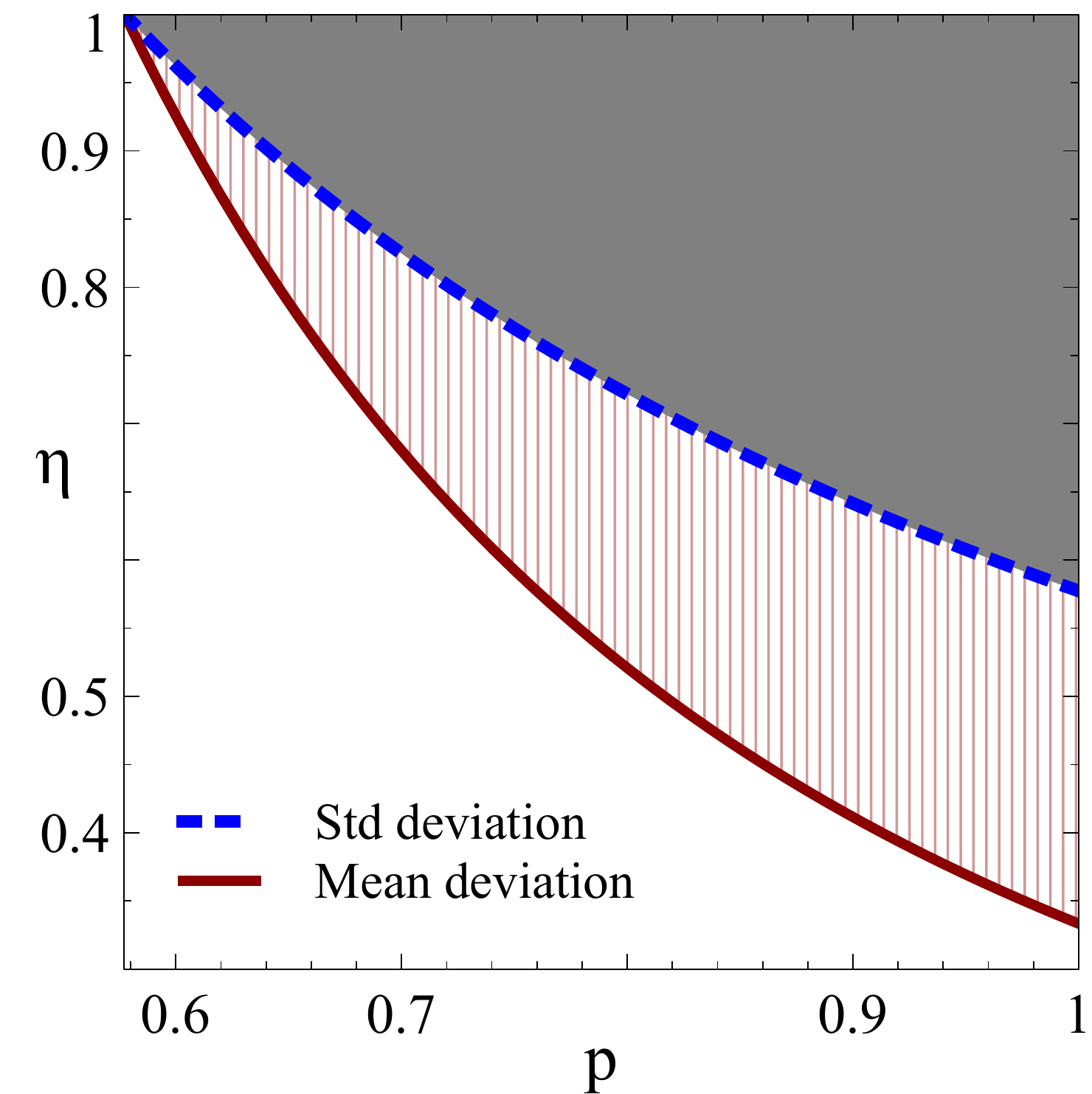}
 \caption{(Color online) Illustrating the power of MD based uncertainty relations in the lossy EPR violation scenario. SD based uncertainty relations cannot detect EPR violation if the detector efficiency $\eta$ plotted along y-axis (with respect to the Werner noise parameter $p$)  is below the dashed blue curve. However, up to the limit of the solid maroon curve, i.e., in the maroon striped zone, the MD based uncertainty relation can still detect such EPR violation.}
 \label{epr_plot}
 \end{figure}
\noindent The full bipartite density matrix can now be represented as $\rho_W \rho_{vac}$, where $\rho_{vac}=\ket{0}\bra{0}$ is the multimode vacuum state in which the undetected particles are collected.  We follow the Schwinger representation to write $\ket{1}$ as $\ket{1,0}$ and $\ket{-1}$ as $\ket{0,1}$, where the $\lbrace i,j\rbrace$ in $\ket{i,j}$ denote the number of particles in spin-up and spin-down configurations respectively. \textcolor{black}{The creation operators at sites A and B are $a_{\pm}^{\dagger}$ and $b_{\pm}^{\dagger}$ respectively, with `+' for statistics of spin-up and `-' for statistics of spin-down particles. Likewise, we denote the creation operators for the vacuum state at sites A and B as $a_{\pm, vac}^{\dagger}$ and $b_{\pm, vac}^{\dagger}$ respectively.} The detection mechanism is described as follows. The particles are led through a beam splitter which couples the field and vacuum modes, after which the modes are transformed as $a_{\pm}\rightarrow \sqrt{\eta}a_{\pm}+\sqrt{1-\eta}a_{\pm,vac}$ and $b_{\pm}\rightarrow \sqrt{\eta}b_{\pm}+\sqrt{1-\eta}b_{\pm,vac}$. The final two-qutrit density matrix $\rho_F$ is now derived by tracing over the lost photon modes. There are a total of nine basis states \begin{itemize}
\item $|u_{1-4}\rangle =\ket{\pm1}_A \ket{\pm1}_B; $
\item $|u_{5,6}\rangle =\ket{\pm1}_A\ket{0}_B ;$
\item $|u_{7,8}\rangle =\ket{0}_A\ket{\pm1}_B$ and,
\item $|u_9\rangle=\ket{0}_A\ket{0}_B.$
\end{itemize}
 The final form of the density matrix $\rho_F$ is now given in block-diagonal form by 
\begin{align*}
\rho_F=\begin{bmatrix}
\eta^2\rho_W & 0 & 0 \\0& \frac{\eta}{2}(1-\eta)\mathbb{I}_4 &0 \\0 &0 & (1-\eta)^2
\end{bmatrix}.
\label{rhoF}
\end{align*}

To find the condition for detecting EPR violation, we use the MD based sum uncertainty relation for the spin operators. Working in the Schwinger representation, we express the spin operators in terms of the particle creation and destruction operators. The spin operators at location A are $J_x^A=(a^{\dagger}_+a_-+a^{\dagger}_-a_+)/2$, $J_y^A=i(a^{\dagger}_-a_+-a^{\dagger}_+a_-)/2$, $J_z^A=(a^{\dagger}_+a_+ -  a^{\dagger}_-a_-)/2$, and the number operator $N^A=a^{\dagger}_+a_++a^{\dagger}_-a_-$. Similarly at location B, the operators $J_x^B,J_y^B,J_z^B$ and $N^B$ are defined using $b_{\pm}$. \textcolor{black}{ For detection of at most a single particle per mode, it can be shown that for a measurement at A, the following inequality holds}.
\begin{align}
\textcolor{black}{\Delta_{M}J_x^A+\Delta_{M}J_y^A+\Delta_{M}J_z^A \geq \frac{3}{2}\braket{N^A}-\frac{\braket{N^A}^2}{2}.}
\end{align}
\noindent To prove the above inequality we have used the inequality $\braket{J_x}^2+\braket{J_y}^2+\braket{J_z}^2\leq \frac{\braket{N}^2}{4}$. The EPR paradox is manifested if the above inequality is violated \textcolor{black}{for inferred mean deviations}, i.e.,
\begin{align}\label{mdddsuminf}
\Delta_{M_{\text{inf}}}J_x^A+\Delta_{M_{\text{inf}}}J_y^A+\Delta_{M_{\text{inf}}}J_z^A < \frac{3}{2}\braket{N^A}-\frac{\braket{N^A}^2}{2}
\end{align}

\textcolor{black}{The inferred mean deviations $\Delta_{M_{\text{inf}}}J_x^A,\Delta_{M_{\text{inf}}}J_y^A$ and $ \Delta_{M_{\text{inf}}}J_z^A$ are the average errors corresponding to the elements of reality that exist for $J_x^A,J_y^A$ and $J_z^A$ respectively. For the loss-included Werner state $\rho_F$, we  compute the inferred mean deviations $\Delta_{M_{\text{inf}}}J_x^A=\Delta_{M_{\text{inf}}}J_y^A = \Delta_{M_{\text{inf}}}J_z^A=\frac{\eta}{2}(1-\eta^{2}p^2)$ and $\braket{N^A}=\eta$.} Putting these values in  \eqref{mdddsuminf}, we get $\eta p^2>\frac{1}{3}$. We compare this result with the corresponding result obtained in \citep{Cavalcanti2009} utilizing the SD based uncertainty relation in Fig. \ref{epr_plot} to note that for the same value of noise parameter $p$, the MD based uncertainty can detect EPR violation with a less efficient detector than the SD based uncertainty. However, for maximum detector efficiency, i.e., for $\eta=1$, we can detect steerability of werner states for $p>\frac{1}{\sqrt{3}}$, which is exactly the same bound derived using standard deviation uncertainty \citep{wernersteerability}.

\emph{Is the mean deviation optimal for detection of EPR violation ?} -We ask at this point, whether any other measure of deviation defined in the same way as the standard deviation or mean deviation, may allow us to detect EPR violation with detectors with even less efficiency. Perhaps, the reader may wonder, it is even possible to detect EPR violation with an extremely inefficient detector, so long as the deviation measure is carefully chosen. In this subsection, we show that such optimism is not correct and mean deviation based bound for EPR violation can not be bettered through choosing a suitable exponent for the measure of deviation, when that exponent is less than unity.

\noindent For an arbitrary $\alpha\leq1$, the generalized  $\alpha$-deviation is defined as
\begin{align}\label{alphadev}
\Delta_M^{\alpha}J= \sum_{a} |\bra{\Psi}\ket{a}|^2 |a-\braket{J}|^{\alpha}.
\end{align}
\noindent where $a$ are the eigenvalues of the observable $J$ and $\ket{a}$ are the corresponding eigenvectors.

When we use $\Delta_M^{\alpha}J$ as the measure of uncertainty for the spin operators they satisfy the following uncertainty relation.
\begin{widetext}
	\begin{align}\label{urel}
	\sum_{i=x,y,z}\Delta_M ^{\alpha}J_i \geq \frac{3\eta}{2^{\alpha}}+\sum_{m=1}^{\infty}\frac{\eta^{2m}\alpha(\alpha-1)...(\alpha-2m-2)}{2^{\alpha}(2m-1)!}\left[\frac{\eta(\alpha-2m-1)}{2m}-1\right].
	\end{align}
\end{widetext}
 
Now if the inferred generalized mean deviations violate the inequality \eqref{urel}, we say that the given state exhibits EPR violation.  The sum of inferred generalized mean deviations are now given as following 
\begin{widetext}
	\begin{align}\label{infdev}
\sum_{i=x,y,z}\Delta_{M_{inf}} ^{\alpha}J_i = \frac{3\eta}{2^{\alpha}}+\sum_{m=1}^{\infty}\frac{3\eta^{2m+1}p^{2m}\alpha(\alpha-1)...(\alpha-2m-2)}{2^{\alpha}(2m-1)!}\left[\frac{\eta(\alpha-2m-1)}{2m}-1\right].
\end{align}
\end{widetext}
\noindent where $p$ is the noise parameter of Werner state.

From Eq.\eqref{urel} and Eq.\eqref{infdev} we note that,  on comparing each term of the series (so that the uncertainty inequality is saturated), we get, from the $m$-th term of the series a relation of the form $\eta\geq \frac{1}{3p^{2m}}$. Therefore the lowest efficiency that we can have is for $m=1$, $\eta=\frac{1}{3p^2}$. This is equal to the lowest efficiency that we get using the mean deviation uncertainty relations.

\subsection{Entanglement detection}
Quantum entanglement is the key resource behind many quantum technologies.  
Coupled with the fact that the complexity of complete state tomography grows exponentially with the dimension, this renders the problem of detection of entanglement in a quntum state via non-tomographic, e.g. witness based methods extremely important. However, the linear witnesses guaranteed to exist vide the Hahn-Banach theorem are often not as strong as desired. Thus considering non-linear witnesses is quite natural. One such family of non-linear witness is furnished by local uncertainty relations, the violation of which implies entanglement in the global state \cite{hofmann}. This method was refined further to derive a necessary criteria for separability in finite-dimensional systems based on inequalities for variances of observables \citep{guehne}. More concretely, it was proven that for an entangled two qubit state $\ket{\Psi_1}=a\ket{00}+b\ket{11}$, with $a>b$ there exist observables $\lbrace \mathcal{O}_i \rbrace$ such that $\sum_i\Delta^2(\mathcal{O}_i)_{\ket{\Psi_1}\bra{\Psi_1}}=0$ ,and the following inequality is obeyed for separable states.
\begin{align}\label{sdsep}
\sum_i\Delta^2(\mathcal{O}_i)\geq 2a^2b^2,
\end{align}
\noindent where $\mathcal{O}_i=\ket{\Psi_i}\bra{\Psi_i}$, $i=1,...,4$ with $\ket{\Psi_2}=a\ket{01}+b\ket{10}$, $\ket{\Psi_3}=b\ket{01}-a\ket{10}$ and $\ket{\Psi_4}=b\ket{00}-a\ket{11}$. Using the above inequality, we can detect the entanglement in members of the Werner family of states $\rho_W^{AB}=p\ket{\Psi_1}\bra{\Psi_1}+\frac{1-p}{4}\mathbb{I}$ for $p> \sqrt{1-\frac{8a^2b^2}{3}}$. If we choose $a=b=\frac{1}{\sqrt{2}},$ we can detect entanglement for $p>\frac{1}{\sqrt{3}}$. Using the MD uncertainty relation, we similarly  obtain the result  $\sum_i\Delta_M(\mathcal{O}_i)_{\ket{\Psi_1}\bra{\Psi_1}}=0$.  For a pure separable state, we note the following inequality being obeyed, i.e.,
\begin{align}\label{mdsep}
\sum_i\Delta_M \mathcal{O}_i\geq 4a^2b^2,
\end{align}
\noindent where $\Delta_M\mathcal{O}_i$ is the mean deviation uncertainty of the observable $\mathcal{O}_i$, for any separable state. Now using  \eqref{mdsep}, it is straightforward to check that we can detect the entanglement of  Werner states for $p> \sqrt{1-\frac{8a^2b^2}{3}}$. Thus, we note that for the Werner family of states, the MD based uncertainty relation is as good a tool as the SD based uncertainty when it comes to entanglement detection. This construction can detect all the bipartite pure entangled states, and, for two qudit systems, many bound entangled states as well \footnote{We also checked that the  MD uncertainty relation based entanglement detection performs as well as the SD uncertainty relation based ones for two qubit Gisin family of  states.}.\\

\textcolor{black}{We note that the scheme of detection of entanglement is quite different than the setup we considered for the  detection of EPR violation earlier. However, in lossy scenario,  it may be shown that one can detect entanglement for any value of detector efficiency using the mean deviation based uncertainty measure considered here. This is consistent with the result obtained in \citep{Cavalcanti2009} which assumes the standard deviation as the measure of uncertainty. EPR steering is, in general, a strictly stronger form of quantum correlation than quantum entanglement  \citep{wiseman,hierarchy,mancini} - thus detection of EPR steering tends to be more demanding than detection of entanglement alone.  }

\section{Conclusion}
\label{V}
In this work, we have provided an alternative formulation of state dependent as well as state independent quantum uncertainty relations in terms of the mean deviation rather than the usual standard deviation based approach. Furthermore, using  F-distribution and Pareto distribution based wave functions, we showed that a definite quantification of quantum uncertainties can be given through our approach which is not possible for standard deviation based approaches in some cases. We have applied the new uncertainty relations in detecting EPR violation in a lossy scenario and in entanglement detection schemes. Of course, applications of mean deviation based uncertainty relations are not confined to these examples. For example, in future, one can formulate new error disturbance relations for successive measurements in terms of mean deviations \citep{ozawa, Branciard1,Branciard2, namrataepl}. Another interesting problem would be to find the analog of the Wigner Yanase skew information \citep{wysi} for the quantum part of the mean deviation based uncertainty \citep{luo} and to study properties thereof, for example, whether this quantity is a true coherence monotone \citep{baumgratz} unlike the WYSI \citep{antigirolami}. For quantum metrological purposes, a mean deviation based formulation of the Cramer Rao bound \citep{paris} may turn out to be useful. Theorists working on quantum gravity have conjectured deformed uncertainty relations \citep{deformed, saurya}. It also remains an interesting direction to explore whether the search for signatures of such deformations, such as the existence of a minimum length scale, can be facilitated by the present work. Another possible direction of work is to explore the identity of intelligent states with respect to arbitrary deviation measures.

\section{Acknowledgement}
The authors acknowledge Department of Atomic Energy, Govt. of India, for providing research fellowships. GS would like to thank Ujjwal Sen and Stephen Gorard for their stimulating inputs. We acknowledge the anonymous referee for suggestions which led to the treatment of EPR violation through generalized $\alpha$-deviations in this paper.

\bibliographystyle{apsrev4-1}
\bibliography{mean_dev_ref}

%merlin.mbs apsrev4-1.bst 2010-07-25 4.21a (PWD, AO, DPC) hacked
%Control: key (0)
%Control: author (72) initials jnrlst
%Control: editor formatted (1) identically to author
%Control: production of article title (-1) disabled
%Control: page (0) single
%Control: year (1) truncated
%Control: production of eprint (0) enabled
\begin{thebibliography}{50}%
\makeatletter
\providecommand \@ifxundefined [1]{%
 \@ifx{#1\undefined}
}%
\providecommand \@ifnum [1]{%
 \ifnum #1\expandafter \@firstoftwo
 \else \expandafter \@secondoftwo
 \fi
}%
\providecommand \@ifx [1]{%
 \ifx #1\expandafter \@firstoftwo
 \else \expandafter \@secondoftwo
 \fi
}%
\providecommand \natexlab [1]{#1}%
\providecommand \enquote  [1]{``#1''}%
\providecommand \bibnamefont  [1]{#1}%
\providecommand \bibfnamefont [1]{#1}%
\providecommand \citenamefont [1]{#1}%
\providecommand \href@noop [0]{\@secondoftwo}%
\providecommand \href [0]{\begingroup \@sanitize@url \@href}%
\providecommand \@href[1]{\@@startlink{#1}\@@href}%
\providecommand \@@href[1]{\endgroup#1\@@endlink}%
\providecommand \@sanitize@url [0]{\catcode `\\12\catcode `\$12\catcode
  `\&12\catcode `\#12\catcode `\^12\catcode `\_12\catcode `\%12\relax}%
\providecommand \@@startlink[1]{}%
\providecommand \@@endlink[0]{}%
\providecommand \url  [0]{\begingroup\@sanitize@url \@url }%
\providecommand \@url [1]{\endgroup\@href {#1}{\urlprefix }}%
\providecommand \urlprefix  [0]{URL }%
\providecommand \Eprint [0]{\href }%
\providecommand \doibase [0]{http://dx.doi.org/}%
\providecommand \selectlanguage [0]{\@gobble}%
\providecommand \bibinfo  [0]{\@secondoftwo}%
\providecommand \bibfield  [0]{\@secondoftwo}%
\providecommand \translation [1]{[#1]}%
\providecommand \BibitemOpen [0]{}%
\providecommand \bibitemStop [0]{}%
\providecommand \bibitemNoStop [0]{.\EOS\space}%
\providecommand \EOS [0]{\spacefactor3000\relax}%
\providecommand \BibitemShut  [1]{\csname bibitem#1\endcsname}%
\let\auto@bib@innerbib\@empty
%</preamble>
\bibitem [{\citenamefont {Heisenberg}(1925)}]{heisenberg}%
  \BibitemOpen
  \bibfield  {author} {\bibinfo {author} {\bibfnamefont {W.}~\bibnamefont
  {Heisenberg}},\ }\href {\doibase 10.1007/BF01328377} {\bibfield  {journal}
  {\bibinfo  {journal} {Zeitschrift f{\"u}r Physik}\ }\textbf {\bibinfo
  {volume} {33}},\ \bibinfo {pages} {879} (\bibinfo {year} {1925})}\BibitemShut
  {NoStop}%
\bibitem [{\citenamefont {{Kennard}}(1927)}]{kennard}%
  \BibitemOpen
  \bibfield  {author} {\bibinfo {author} {\bibfnamefont {E.~H.}\ \bibnamefont
  {{Kennard}}},\ }\href {\doibase 10.1007/BF01391200} {\bibfield  {journal}
  {\bibinfo  {journal} {Zeitschrift fur Physik}\ }\textbf {\bibinfo {volume}
  {44}},\ \bibinfo {pages} {326} (\bibinfo {year} {1927})}\BibitemShut
  {NoStop}%
\bibitem [{\citenamefont {Robertson}(1929)}]{robertson}%
  \BibitemOpen
  \bibfield  {author} {\bibinfo {author} {\bibfnamefont {H.~P.}\ \bibnamefont
  {Robertson}},\ }\href {\doibase 10.1103/PhysRev.34.163} {\bibfield  {journal}
  {\bibinfo  {journal} {Phys. Rev.}\ }\textbf {\bibinfo {volume} {34}},\
  \bibinfo {pages} {163} (\bibinfo {year} {1929})}\BibitemShut {NoStop}%
\bibitem [{\citenamefont {Fisher}(1925)}]{fisherbook}%
  \BibitemOpen
  \bibfield  {author} {\bibinfo {author} {\bibfnamefont {R.}~\bibnamefont
  {Fisher}},\ }\href {https://books.google.co.in/books?id=4bTttAJR5kEC} {\emph
  {\bibinfo {title} {Statistical Methods For Research Workers}}},\ Cosmo study
  guides\ (\bibinfo  {publisher} {Cosmo Publications},\ \bibinfo {year}
  {1925})\BibitemShut {NoStop}%
\bibitem [{\citenamefont {Feller}(1968)}]{Feller-book}%
  \BibitemOpen
  \bibfield  {author} {\bibinfo {author} {\bibfnamefont {W.}~\bibnamefont
  {Feller}},\ }\href@noop {} {\emph {\bibinfo {title} {An Introduction to
  Probability Theory and its Applications}}}\ (\bibinfo  {publisher} {Wiley},\
  \bibinfo {year} {1968})\BibitemShut {NoStop}%
\bibitem [{\citenamefont {Coles}\ \emph {et~al.}(2017)\citenamefont {Coles},
  \citenamefont {Berta}, \citenamefont {Tomamichel},\ and\ \citenamefont
  {Wehner}}]{eur}%
  \BibitemOpen
  \bibfield  {author} {\bibinfo {author} {\bibfnamefont {P.~J.}\ \bibnamefont
  {Coles}}, \bibinfo {author} {\bibfnamefont {M.}~\bibnamefont {Berta}},
  \bibinfo {author} {\bibfnamefont {M.}~\bibnamefont {Tomamichel}}, \ and\
  \bibinfo {author} {\bibfnamefont {S.}~\bibnamefont {Wehner}},\ }\href
  {\doibase 10.1103/RevModPhys.89.015002} {\bibfield  {journal} {\bibinfo
  {journal} {Rev. Mod. Phys.}\ }\textbf {\bibinfo {volume} {89}},\ \bibinfo
  {pages} {015002} (\bibinfo {year} {2017})}\BibitemShut {NoStop}%
\bibitem [{\citenamefont {Sakurai}(1993)}]{sakurai-book}%
  \BibitemOpen
  \bibfield  {author} {\bibinfo {author} {\bibfnamefont {J.~J.}\ \bibnamefont
  {Sakurai}},\ }\href@noop {} {\emph {\bibinfo {title} {Modern Quantum
  Mechanics}}}\ (\bibinfo  {publisher} {Addison Wesley},\ \bibinfo {year}
  {1993})\BibitemShut {NoStop}%
\bibitem [{\citenamefont {Maccone}\ and\ \citenamefont
  {Pati}(2014)}]{macconepati}%
  \BibitemOpen
  \bibfield  {author} {\bibinfo {author} {\bibfnamefont {L.}~\bibnamefont
  {Maccone}}\ and\ \bibinfo {author} {\bibfnamefont {A.~K.}\ \bibnamefont
  {Pati}},\ }\href {\doibase 10.1103/PhysRevLett.113.260401} {\bibfield
  {journal} {\bibinfo  {journal} {Phys. Rev. Lett.}\ }\textbf {\bibinfo
  {volume} {113}},\ \bibinfo {pages} {260401} (\bibinfo {year}
  {2014})}\BibitemShut {NoStop}%
\bibitem [{\citenamefont {{Mukhopadhyay}}\ and\ \citenamefont
  {{Pati}}(2018)}]{newurpaper}%
  \BibitemOpen
  \bibfield  {author} {\bibinfo {author} {\bibfnamefont {C.}~\bibnamefont
  {{Mukhopadhyay}}}\ and\ \bibinfo {author} {\bibfnamefont {A.~K.}\
  \bibnamefont {{Pati}}},\ }\href@noop {} {\bibfield  {journal} {\bibinfo
  {journal} {ArXiv e-prints}\ } (\bibinfo {year} {2018})},\ \Eprint
  {http://arxiv.org/abs/1806.11347} {arXiv:1806.11347 [quant-ph]} \BibitemShut
  {NoStop}%
\bibitem [{\citenamefont {{Song}}\ \emph {et~al.}(2017)\citenamefont {{Song}},
  \citenamefont {{Li}}, \citenamefont {{Peng}},\ and\ \citenamefont
  {{Qiao}}}]{mult}%
  \BibitemOpen
  \bibfield  {author} {\bibinfo {author} {\bibfnamefont {Q.-C.}\ \bibnamefont
  {{Song}}}, \bibinfo {author} {\bibfnamefont {J.-L.}\ \bibnamefont {{Li}}},
  \bibinfo {author} {\bibfnamefont {G.-X.}\ \bibnamefont {{Peng}}}, \ and\
  \bibinfo {author} {\bibfnamefont {C.-F.}\ \bibnamefont {{Qiao}}},\ }\href
  {\doibase 10.1038/srep44764} {\bibfield  {journal} {\bibinfo  {journal}
  {Scientific Reports}\ }\textbf {\bibinfo {volume} {7}},\ \bibinfo {eid}
  {44764} (\bibinfo {year} {2017})}\BibitemShut {NoStop}%
\bibitem [{\citenamefont {Mondal}\ \emph {et~al.}(2017)\citenamefont {Mondal},
  \citenamefont {Bagchi},\ and\ \citenamefont {Pati}}]{shrobona}%
  \BibitemOpen
  \bibfield  {author} {\bibinfo {author} {\bibfnamefont {D.}~\bibnamefont
  {Mondal}}, \bibinfo {author} {\bibfnamefont {S.}~\bibnamefont {Bagchi}}, \
  and\ \bibinfo {author} {\bibfnamefont {A.~K.}\ \bibnamefont {Pati}},\ }\href
  {\doibase 10.1103/PhysRevA.95.052117} {\bibfield  {journal} {\bibinfo
  {journal} {Phys. Rev. A}\ }\textbf {\bibinfo {volume} {95}},\ \bibinfo
  {pages} {052117} (\bibinfo {year} {2017})}\BibitemShut {NoStop}%
\bibitem [{\citenamefont {{Bera}}\ \emph {et~al.}(2017)\citenamefont {{Bera}},
  \citenamefont {{Das}}, \citenamefont {{Sen(De)}},\ and\ \citenamefont
  {{Sen}}}]{bera_median}%
  \BibitemOpen
  \bibfield  {author} {\bibinfo {author} {\bibfnamefont {A.}~\bibnamefont
  {{Bera}}}, \bibinfo {author} {\bibfnamefont {D.}~\bibnamefont {{Das}}},
  \bibinfo {author} {\bibfnamefont {A.}~\bibnamefont {{Sen(De)}}}, \ and\
  \bibinfo {author} {\bibfnamefont {U.}~\bibnamefont {{Sen}}},\ }\href@noop {}
  {\  (\bibinfo {year} {2017})},\ \Eprint {http://arxiv.org/abs/1706.00720}
  {arXiv:1706.00720} \BibitemShut {NoStop}%
\bibitem [{\citenamefont {Bannerjee}(2008)}]{giri}%
  \BibitemOpen
  \bibfield  {author} {\bibinfo {author} {\bibfnamefont {P.}~\bibnamefont
  {Bannerjee}},\ }\href {https://books.google.co.in/books?id=E1zn5ukZemIC}
  {\emph {\bibinfo {title} {Introduction To Statistics}}}\ (\bibinfo
  {publisher} {Academic Publishers, Kolkata},\ \bibinfo {year}
  {2008})\BibitemShut {NoStop}%
\bibitem [{\citenamefont {Gorard}(2005)}]{gorard}%
  \BibitemOpen
  \bibfield  {author} {\bibinfo {author} {\bibfnamefont {S.}~\bibnamefont
  {Gorard}},\ }\href {http://www.jstor.org/stable/3699276} {\bibfield
  {journal} {\bibinfo  {journal} {British Journal of Educational Studies}\
  }\textbf {\bibinfo {volume} {53}},\ \bibinfo {pages} {417} (\bibinfo {year}
  {2005})}\BibitemShut {NoStop}%
\bibitem [{Note1()}]{Note1}%
  \BibitemOpen
  \bibinfo {note} {Indeed, both SD and MD can be expressed as distance measures
  in this view - SD as the Euclidean distance, MD as the
  $l_1$-distance.}\BibitemShut {Stop}%
\bibitem [{\citenamefont {Berta}\ \emph {et~al.}(2010)\citenamefont {Berta},
  \citenamefont {Christandl}, \citenamefont {Colbeck}, \citenamefont {Renes},\
  and\ \citenamefont {Renner}}]{berta-nature}%
  \BibitemOpen
  \bibfield  {author} {\bibinfo {author} {\bibfnamefont {M.}~\bibnamefont
  {Berta}}, \bibinfo {author} {\bibfnamefont {M.}~\bibnamefont {Christandl}},
  \bibinfo {author} {\bibfnamefont {R.}~\bibnamefont {Colbeck}}, \bibinfo
  {author} {\bibfnamefont {J.~M.}\ \bibnamefont {Renes}}, \ and\ \bibinfo
  {author} {\bibfnamefont {R.}~\bibnamefont {Renner}},\ }\href
  {http://dx.doi.org/10.1038/nphys1734} {\bibfield  {journal} {\bibinfo
  {journal} {Nature Physics}\ }\textbf {\bibinfo {volume} {6}},\ \bibinfo
  {pages} {659 EP } (\bibinfo {year} {2010})}\BibitemShut {NoStop}%
\bibitem [{\citenamefont {{Wehner}}\ and\ \citenamefont
  {{Winter}}(2010)}]{eurwehner}%
  \BibitemOpen
  \bibfield  {author} {\bibinfo {author} {\bibfnamefont {S.}~\bibnamefont
  {{Wehner}}}\ and\ \bibinfo {author} {\bibfnamefont {A.}~\bibnamefont
  {{Winter}}},\ }\href {\doibase 10.1088/1367-2630/12/2/025009} {\bibfield
  {journal} {\bibinfo  {journal} {New Journal of Physics}\ }\textbf {\bibinfo
  {volume} {12}},\ \bibinfo {eid} {025009} (\bibinfo {year}
  {2010})}\BibitemShut {NoStop}%
\bibitem [{\citenamefont {Cavalcanti}\ and\ \citenamefont
  {Reid}(2007)}]{cavalcanti}%
  \BibitemOpen
  \bibfield  {author} {\bibinfo {author} {\bibfnamefont {E.~G.}\ \bibnamefont
  {Cavalcanti}}\ and\ \bibinfo {author} {\bibfnamefont {M.~D.}\ \bibnamefont
  {Reid}},\ }\href {\doibase 10.1080/09500340701639623} {\bibfield  {journal}
  {\bibinfo  {journal} {Journal of Modern Optics}\ }\textbf {\bibinfo {volume}
  {54}},\ \bibinfo {pages} {2373} (\bibinfo {year} {2007})},\ \Eprint
  {http://arxiv.org/abs/https://doi.org/10.1080/09500340701639623}
  {https://doi.org/10.1080/09500340701639623} \BibitemShut {NoStop}%
\bibitem [{\citenamefont {Hofmann}\ and\ \citenamefont
  {Takeuchi}(2003)}]{hofmann}%
  \BibitemOpen
  \bibfield  {author} {\bibinfo {author} {\bibfnamefont {H.~F.}\ \bibnamefont
  {Hofmann}}\ and\ \bibinfo {author} {\bibfnamefont {S.}~\bibnamefont
  {Takeuchi}},\ }\href {\doibase 10.1103/PhysRevA.68.032103} {\bibfield
  {journal} {\bibinfo  {journal} {Phys. Rev. A}\ }\textbf {\bibinfo {volume}
  {68}},\ \bibinfo {pages} {032103} (\bibinfo {year} {2003})}\BibitemShut
  {NoStop}%
\bibitem [{\citenamefont {G\"uhne}(2004)}]{guehne}%
  \BibitemOpen
  \bibfield  {author} {\bibinfo {author} {\bibfnamefont {O.}~\bibnamefont
  {G\"uhne}},\ }\href {\doibase 10.1103/PhysRevLett.92.117903} {\bibfield
  {journal} {\bibinfo  {journal} {Phys. Rev. Lett.}\ }\textbf {\bibinfo
  {volume} {92}},\ \bibinfo {pages} {117903} (\bibinfo {year}
  {2004})}\BibitemShut {NoStop}%
\bibitem [{\citenamefont {{Pati}}(1999)}]{pati}%
  \BibitemOpen
  \bibfield  {author} {\bibinfo {author} {\bibfnamefont {A.~K.}\ \bibnamefont
  {{Pati}}},\ }\href {\doibase 10.1016/S0375-9601(99)00701-X} {\bibfield
  {journal} {\bibinfo  {journal} {Physics Letters A}\ }\textbf {\bibinfo
  {volume} {262}},\ \bibinfo {pages} {296} (\bibinfo {year} {1999})},\ \Eprint
  {http://arxiv.org/abs/quant-ph/9901033} {quant-ph/9901033} \BibitemShut
  {NoStop}%
\bibitem [{\citenamefont {Ballentine}(1998)}]{Ballentine-book}%
  \BibitemOpen
  \bibfield  {author} {\bibinfo {author} {\bibfnamefont {L.}~\bibnamefont
  {Ballentine}},\ }\href@noop {} {\emph {\bibinfo {title} {Quantum Mechanics: A
  Modern Development}}}\ (\bibinfo  {publisher} {World Scientific},\ \bibinfo
  {year} {1998})\BibitemShut {NoStop}%
\bibitem [{\citenamefont {Grosshans}()}]{stack}%
  \BibitemOpen
  \bibfield  {author} {\bibinfo {author} {\bibfnamefont {F.}~\bibnamefont
  {Grosshans}},\ }\href {https://physics.stackexchange.com/q/251858} {\enquote
  {\bibinfo {title} {Heisenberg's uncertainty principle for mean deviation?}}\
  }\bibinfo {howpublished} {Physics Stack Exchange},\ \bibinfo {note}
  {https://physics.stackexchange.com/q/251858 (version: 2017-04-13)},\ \Eprint
  {http://arxiv.org/abs/https://physics.stackexchange.com/q/251858}
  {https://physics.stackexchange.com/q/251858} \BibitemShut {NoStop}%
\bibitem [{\citenamefont {Huang}(2012)}]{huang}%
  \BibitemOpen
  \bibfield  {author} {\bibinfo {author} {\bibfnamefont {Y.}~\bibnamefont
  {Huang}},\ }\href {\doibase 10.1103/PhysRevA.86.024101} {\bibfield  {journal}
  {\bibinfo  {journal} {Phys. Rev. A}\ }\textbf {\bibinfo {volume} {86}},\
  \bibinfo {pages} {024101} (\bibinfo {year} {2012})}\BibitemShut {NoStop}%
\bibitem [{\citenamefont {Maassen}\ and\ \citenamefont
  {Uffink}(1988)}]{maassenuffink}%
  \BibitemOpen
  \bibfield  {author} {\bibinfo {author} {\bibfnamefont {H.}~\bibnamefont
  {Maassen}}\ and\ \bibinfo {author} {\bibfnamefont {J.~B.~M.}\ \bibnamefont
  {Uffink}},\ }\href {\doibase 10.1103/PhysRevLett.60.1103} {\bibfield
  {journal} {\bibinfo  {journal} {Phys. Rev. Lett.}\ }\textbf {\bibinfo
  {volume} {60}},\ \bibinfo {pages} {1103} (\bibinfo {year}
  {1988})}\BibitemShut {NoStop}%
\bibitem [{\citenamefont {{Mandelshtam}}\ and\ \citenamefont
  {{Tamm}}(1945)}]{mandelstam}%
  \BibitemOpen
  \bibfield  {author} {\bibinfo {author} {\bibfnamefont {L.~I.}\ \bibnamefont
  {{Mandelshtam}}}\ and\ \bibinfo {author} {\bibfnamefont {I.~E.}\ \bibnamefont
  {{Tamm}}},\ }\href@noop {} {\bibfield  {journal} {\bibinfo  {journal} {J.
  Phys. (USSR)}\ }\textbf {\bibinfo {volume} {9}},\ \bibinfo {eid} {249}
  (\bibinfo {year} {1945})}\BibitemShut {NoStop}%
\bibitem [{\citenamefont {Anandan}\ and\ \citenamefont
  {Aharonov}(1990)}]{anandan}%
  \BibitemOpen
  \bibfield  {author} {\bibinfo {author} {\bibfnamefont {J.}~\bibnamefont
  {Anandan}}\ and\ \bibinfo {author} {\bibfnamefont {Y.}~\bibnamefont
  {Aharonov}},\ }\href {\doibase 10.1103/PhysRevLett.65.1697} {\bibfield
  {journal} {\bibinfo  {journal} {Phys. Rev. Lett.}\ }\textbf {\bibinfo
  {volume} {65}},\ \bibinfo {pages} {1697} (\bibinfo {year}
  {1990})}\BibitemShut {NoStop}%
\bibitem [{\citenamefont {{Vaidman}}(1992)}]{vaidman}%
  \BibitemOpen
  \bibfield  {author} {\bibinfo {author} {\bibfnamefont {L.}~\bibnamefont
  {{Vaidman}}},\ }\href {\doibase 10.1119/1.16940} {\bibfield  {journal}
  {\bibinfo  {journal} {American Journal of Physics}\ }\textbf {\bibinfo
  {volume} {60}},\ \bibinfo {pages} {182} (\bibinfo {year} {1992})}\BibitemShut
  {NoStop}%
\bibitem [{\citenamefont {Mal}\ \emph {et~al.}(2013)\citenamefont {Mal},
  \citenamefont {Pramanik},\ and\ \citenamefont {Majumdar}}]{mal}%
  \BibitemOpen
  \bibfield  {author} {\bibinfo {author} {\bibfnamefont {S.}~\bibnamefont
  {Mal}}, \bibinfo {author} {\bibfnamefont {T.}~\bibnamefont {Pramanik}}, \
  and\ \bibinfo {author} {\bibfnamefont {A.~S.}\ \bibnamefont {Majumdar}},\
  }\href {\doibase 10.1103/PhysRevA.87.012105} {\bibfield  {journal} {\bibinfo
  {journal} {Phys. Rev. A}\ }\textbf {\bibinfo {volume} {87}},\ \bibinfo
  {pages} {012105} (\bibinfo {year} {2013})}\BibitemShut {NoStop}%
\bibitem [{\citenamefont {Reid}(1989)}]{reid}%
  \BibitemOpen
  \bibfield  {author} {\bibinfo {author} {\bibfnamefont {M.~D.}\ \bibnamefont
  {Reid}},\ }\href {\doibase 10.1103/PhysRevA.40.913} {\bibfield  {journal}
  {\bibinfo  {journal} {Phys. Rev. A}\ }\textbf {\bibinfo {volume} {40}},\
  \bibinfo {pages} {913} (\bibinfo {year} {1989})}\BibitemShut {NoStop}%
\bibitem [{\citenamefont {Oppenheim}\ and\ \citenamefont
  {Wehner}(2010)}]{wehner}%
  \BibitemOpen
  \bibfield  {author} {\bibinfo {author} {\bibfnamefont {J.}~\bibnamefont
  {Oppenheim}}\ and\ \bibinfo {author} {\bibfnamefont {S.}~\bibnamefont
  {Wehner}},\ }\href {\doibase 10.1126/science.1192065} {\bibfield  {journal}
  {\bibinfo  {journal} {Science}\ }\textbf {\bibinfo {volume} {330}},\ \bibinfo
  {pages} {1072} (\bibinfo {year} {2010})},\ \Eprint
  {http://arxiv.org/abs/http://science.sciencemag.org/content/330/6007/1072.full.pdf}
  {http://science.sciencemag.org/content/330/6007/1072.full.pdf} \BibitemShut
  {NoStop}%
\bibitem [{\citenamefont {Einstein}\ \emph {et~al.}(1935)\citenamefont
  {Einstein}, \citenamefont {Podolsky},\ and\ \citenamefont {Rosen}}]{epr}%
  \BibitemOpen
  \bibfield  {author} {\bibinfo {author} {\bibfnamefont {A.}~\bibnamefont
  {Einstein}}, \bibinfo {author} {\bibfnamefont {B.}~\bibnamefont {Podolsky}},
  \ and\ \bibinfo {author} {\bibfnamefont {N.}~\bibnamefont {Rosen}},\ }\href
  {\doibase 10.1103/PhysRev.47.777} {\bibfield  {journal} {\bibinfo  {journal}
  {Phys. Rev.}\ }\textbf {\bibinfo {volume} {47}},\ \bibinfo {pages} {777}
  (\bibinfo {year} {1935})}\BibitemShut {NoStop}%
\bibitem [{\citenamefont {Bell}(2004)}]{bell}%
  \BibitemOpen
  \bibfield  {author} {\bibinfo {author} {\bibfnamefont {J.}~\bibnamefont
  {Bell}},\ }\href {https://books.google.co.in/books?id=FGnnHxh2YtQC} {\emph
  {\bibinfo {title} {Speakable and Unspeakable in Quantum Mechanics: Collected
  Papers on Quantum Philosophy}}},\ Collected papers on quantum philosophy\
  (\bibinfo  {publisher} {Cambridge University Press},\ \bibinfo {year}
  {2004})\BibitemShut {NoStop}%
\bibitem [{\citenamefont {Wiseman}\ \emph {et~al.}(2007)\citenamefont
  {Wiseman}, \citenamefont {Jones},\ and\ \citenamefont {Doherty}}]{wiseman}%
  \BibitemOpen
  \bibfield  {author} {\bibinfo {author} {\bibfnamefont {H.~M.}\ \bibnamefont
  {Wiseman}}, \bibinfo {author} {\bibfnamefont {S.~J.}\ \bibnamefont {Jones}},
  \ and\ \bibinfo {author} {\bibfnamefont {A.~C.}\ \bibnamefont {Doherty}},\
  }\href {\doibase 10.1103/PhysRevLett.98.140402} {\bibfield  {journal}
  {\bibinfo  {journal} {Phys. Rev. Lett.}\ }\textbf {\bibinfo {volume} {98}},\
  \bibinfo {pages} {140402} (\bibinfo {year} {2007})}\BibitemShut {NoStop}%
\bibitem [{\citenamefont {Cavalcanti}\ \emph
  {et~al.}(2009{\natexlab{a}})\citenamefont {Cavalcanti}, \citenamefont
  {Drummond}, \citenamefont {Bachor},\ and\ \citenamefont
  {Reid}}]{Cavalcanti2009}%
  \BibitemOpen
  \bibfield  {author} {\bibinfo {author} {\bibfnamefont {E.~G.}\ \bibnamefont
  {Cavalcanti}}, \bibinfo {author} {\bibfnamefont {P.~D.}\ \bibnamefont
  {Drummond}}, \bibinfo {author} {\bibfnamefont {H.~A.}\ \bibnamefont
  {Bachor}}, \ and\ \bibinfo {author} {\bibfnamefont {M.~D.}\ \bibnamefont
  {Reid}},\ }\href {\doibase 10.1364/OE.17.018693} {\bibfield  {journal}
  {\bibinfo  {journal} {Opt. Express}\ }\textbf {\bibinfo {volume} {17}},\
  \bibinfo {pages} {18693} (\bibinfo {year} {2009}{\natexlab{a}})}\BibitemShut
  {NoStop}%
\bibitem [{\citenamefont {Cavalcanti}\ \emph
  {et~al.}(2009{\natexlab{b}})\citenamefont {Cavalcanti}, \citenamefont
  {Jones}, \citenamefont {Wiseman},\ and\ \citenamefont
  {Reid}}]{wernersteerability}%
  \BibitemOpen
  \bibfield  {author} {\bibinfo {author} {\bibfnamefont {E.~G.}\ \bibnamefont
  {Cavalcanti}}, \bibinfo {author} {\bibfnamefont {S.~J.}\ \bibnamefont
  {Jones}}, \bibinfo {author} {\bibfnamefont {H.~M.}\ \bibnamefont {Wiseman}},
  \ and\ \bibinfo {author} {\bibfnamefont {M.~D.}\ \bibnamefont {Reid}},\
  }\href {\doibase 10.1103/PhysRevA.80.032112} {\bibfield  {journal} {\bibinfo
  {journal} {Phys. Rev. A}\ }\textbf {\bibinfo {volume} {80}},\ \bibinfo
  {pages} {032112} (\bibinfo {year} {2009}{\natexlab{b}})}\BibitemShut
  {NoStop}%
\bibitem [{Note2()}]{Note2}%
  \BibitemOpen
  \bibinfo {note} {We also checked that the MD uncertainty relation based
  entanglement detection performs as well as the SD uncertainty relation based
  ones for two qubit Gisin family of states.}\BibitemShut {Stop}%
\bibitem [{\citenamefont {Quintino}\ \emph {et~al.}(2015)\citenamefont
  {Quintino}, \citenamefont {V\'ertesi}, \citenamefont {Cavalcanti},
  \citenamefont {Augusiak}, \citenamefont {Demianowicz}, \citenamefont
  {Ac\'{\i}n},\ and\ \citenamefont {Brunner}}]{hierarchy}%
  \BibitemOpen
  \bibfield  {author} {\bibinfo {author} {\bibfnamefont {M.~T.}\ \bibnamefont
  {Quintino}}, \bibinfo {author} {\bibfnamefont {T.}~\bibnamefont {V\'ertesi}},
  \bibinfo {author} {\bibfnamefont {D.}~\bibnamefont {Cavalcanti}}, \bibinfo
  {author} {\bibfnamefont {R.}~\bibnamefont {Augusiak}}, \bibinfo {author}
  {\bibfnamefont {M.}~\bibnamefont {Demianowicz}}, \bibinfo {author}
  {\bibfnamefont {A.}~\bibnamefont {Ac\'{\i}n}}, \ and\ \bibinfo {author}
  {\bibfnamefont {N.}~\bibnamefont {Brunner}},\ }\href {\doibase
  10.1103/PhysRevA.92.032107} {\bibfield  {journal} {\bibinfo  {journal} {Phys.
  Rev. A}\ }\textbf {\bibinfo {volume} {92}},\ \bibinfo {pages} {032107}
  (\bibinfo {year} {2015})}\BibitemShut {NoStop}%
\bibitem [{\citenamefont {Mancini}\ \emph {et~al.}(2002)\citenamefont
  {Mancini}, \citenamefont {Giovannetti}, \citenamefont {Vitali},\ and\
  \citenamefont {Tombesi}}]{mancini}%
  \BibitemOpen
  \bibfield  {author} {\bibinfo {author} {\bibfnamefont {S.}~\bibnamefont
  {Mancini}}, \bibinfo {author} {\bibfnamefont {V.}~\bibnamefont
  {Giovannetti}}, \bibinfo {author} {\bibfnamefont {D.}~\bibnamefont {Vitali}},
  \ and\ \bibinfo {author} {\bibfnamefont {P.}~\bibnamefont {Tombesi}},\ }\href
  {\doibase 10.1103/PhysRevLett.88.120401} {\bibfield  {journal} {\bibinfo
  {journal} {Phys. Rev. Lett.}\ }\textbf {\bibinfo {volume} {88}},\ \bibinfo
  {pages} {120401} (\bibinfo {year} {2002})}\BibitemShut {NoStop}%
\bibitem [{\citenamefont {Ozawa}(2003)}]{ozawa}%
  \BibitemOpen
  \bibfield  {author} {\bibinfo {author} {\bibfnamefont {M.}~\bibnamefont
  {Ozawa}},\ }\href {\doibase 10.1103/PhysRevA.67.042105} {\bibfield  {journal}
  {\bibinfo  {journal} {Phys. Rev. A}\ }\textbf {\bibinfo {volume} {67}},\
  \bibinfo {pages} {042105} (\bibinfo {year} {2003})}\BibitemShut {NoStop}%
\bibitem [{\citenamefont {Branciard}(2014)}]{Branciard1}%
  \BibitemOpen
  \bibfield  {author} {\bibinfo {author} {\bibfnamefont {C.}~\bibnamefont
  {Branciard}},\ }\href {\doibase 10.1103/PhysRevA.89.022124} {\bibfield
  {journal} {\bibinfo  {journal} {Phys. Rev. A}\ }\textbf {\bibinfo {volume}
  {89}},\ \bibinfo {pages} {022124} (\bibinfo {year} {2014})}\BibitemShut
  {NoStop}%
\bibitem [{\citenamefont {Branciard}(2013)}]{Branciard2}%
  \BibitemOpen
  \bibfield  {author} {\bibinfo {author} {\bibfnamefont {C.}~\bibnamefont
  {Branciard}},\ }\href {\doibase 10.1073/pnas.1219331110} {\bibfield
  {journal} {\bibinfo  {journal} {Proceedings of the National Academy of
  Sciences}\ }\textbf {\bibinfo {volume} {110}},\ \bibinfo {pages} {6742}
  (\bibinfo {year} {2013})},\ \Eprint
  {http://arxiv.org/abs/http://www.pnas.org/content/110/17/6742.full.pdf}
  {http://www.pnas.org/content/110/17/6742.full.pdf} \BibitemShut {NoStop}%
\bibitem [{\citenamefont {Mukhopadhyay}\ \emph {et~al.}(2016)\citenamefont
  {Mukhopadhyay}, \citenamefont {Shukla},\ and\ \citenamefont
  {Pati}}]{namrataepl}%
  \BibitemOpen
  \bibfield  {author} {\bibinfo {author} {\bibfnamefont {C.}~\bibnamefont
  {Mukhopadhyay}}, \bibinfo {author} {\bibfnamefont {N.}~\bibnamefont
  {Shukla}}, \ and\ \bibinfo {author} {\bibfnamefont {A.~K.}\ \bibnamefont
  {Pati}},\ }\href {http://stacks.iop.org/0295-5075/113/i=5/a=50002} {\bibfield
   {journal} {\bibinfo  {journal} {EPL (Europhysics Letters)}\ }\textbf
  {\bibinfo {volume} {113}},\ \bibinfo {pages} {50002} (\bibinfo {year}
  {2016})}\BibitemShut {NoStop}%
\bibitem [{\citenamefont {Wigner}\ and\ \citenamefont {Yanase}(1963)}]{wysi}%
  \BibitemOpen
  \bibfield  {author} {\bibinfo {author} {\bibfnamefont {E.~P.}\ \bibnamefont
  {Wigner}}\ and\ \bibinfo {author} {\bibfnamefont {M.~M.}\ \bibnamefont
  {Yanase}},\ }\href {http://www.pnas.org/content/49/6/910.short} {\bibfield
  {journal} {\bibinfo  {journal} {Proceedings of the National Academy of
  Sciences}\ }\textbf {\bibinfo {volume} {49}},\ \bibinfo {pages} {910}
  (\bibinfo {year} {1963})},\ \Eprint
  {http://arxiv.org/abs/http://www.pnas.org/content/49/6/910.full.pdf}
  {http://www.pnas.org/content/49/6/910.full.pdf} \BibitemShut {NoStop}%
\bibitem [{\citenamefont {Luo}(2003)}]{luo}%
  \BibitemOpen
  \bibfield  {author} {\bibinfo {author} {\bibfnamefont {S.}~\bibnamefont
  {Luo}},\ }\href {\doibase 10.1103/PhysRevLett.91.180403} {\bibfield
  {journal} {\bibinfo  {journal} {Phys. Rev. Lett.}\ }\textbf {\bibinfo
  {volume} {91}},\ \bibinfo {pages} {180403} (\bibinfo {year}
  {2003})}\BibitemShut {NoStop}%
\bibitem [{\citenamefont {Baumgratz}\ \emph {et~al.}(2014)\citenamefont
  {Baumgratz}, \citenamefont {Cramer},\ and\ \citenamefont
  {Plenio}}]{baumgratz}%
  \BibitemOpen
  \bibfield  {author} {\bibinfo {author} {\bibfnamefont {T.}~\bibnamefont
  {Baumgratz}}, \bibinfo {author} {\bibfnamefont {M.}~\bibnamefont {Cramer}}, \
  and\ \bibinfo {author} {\bibfnamefont {M.~B.}\ \bibnamefont {Plenio}},\
  }\href {\doibase 10.1103/PhysRevLett.113.140401} {\bibfield  {journal}
  {\bibinfo  {journal} {Phys. Rev. Lett.}\ }\textbf {\bibinfo {volume} {113}},\
  \bibinfo {pages} {140401} (\bibinfo {year} {2014})}\BibitemShut {NoStop}%
\bibitem [{\citenamefont {Du}\ and\ \citenamefont {Bai}(2015)}]{antigirolami}%
  \BibitemOpen
  \bibfield  {author} {\bibinfo {author} {\bibfnamefont {S.}~\bibnamefont
  {Du}}\ and\ \bibinfo {author} {\bibfnamefont {Z.}~\bibnamefont {Bai}},\
  }\href {\doibase https://doi.org/10.1016/j.aop.2015.04.023} {\bibfield
  {journal} {\bibinfo  {journal} {Annals of Physics}\ }\textbf {\bibinfo
  {volume} {359}},\ \bibinfo {pages} {136 } (\bibinfo {year}
  {2015})}\BibitemShut {NoStop}%
\bibitem [{\citenamefont {{Paris}}(2008)}]{paris}%
  \BibitemOpen
  \bibfield  {author} {\bibinfo {author} {\bibfnamefont {M.~G.~A.}\
  \bibnamefont {{Paris}}},\ }\href@noop {} {\bibfield  {journal} {\bibinfo
  {journal} {ArXiv e-prints}\ } (\bibinfo {year} {2008})},\ \Eprint
  {http://arxiv.org/abs/0804.2981} {arXiv:0804.2981 [quant-ph]} \BibitemShut
  {NoStop}%
\bibitem [{\citenamefont {Kempf}\ \emph {et~al.}(1995)\citenamefont {Kempf},
  \citenamefont {Mangano},\ and\ \citenamefont {Mann}}]{deformed}%
  \BibitemOpen
  \bibfield  {author} {\bibinfo {author} {\bibfnamefont {A.}~\bibnamefont
  {Kempf}}, \bibinfo {author} {\bibfnamefont {G.}~\bibnamefont {Mangano}}, \
  and\ \bibinfo {author} {\bibfnamefont {R.~B.}\ \bibnamefont {Mann}},\ }\href
  {\doibase 10.1103/PhysRevD.52.1108} {\bibfield  {journal} {\bibinfo
  {journal} {Phys. Rev. D}\ }\textbf {\bibinfo {volume} {52}},\ \bibinfo
  {pages} {1108} (\bibinfo {year} {1995})}\BibitemShut {NoStop}%
\bibitem [{\citenamefont {Das}\ and\ \citenamefont {Vagenas}(2008)}]{saurya}%
  \BibitemOpen
  \bibfield  {author} {\bibinfo {author} {\bibfnamefont {S.}~\bibnamefont
  {Das}}\ and\ \bibinfo {author} {\bibfnamefont {E.~C.}\ \bibnamefont
  {Vagenas}},\ }\href {\doibase 10.1103/PhysRevLett.101.221301} {\bibfield
  {journal} {\bibinfo  {journal} {Phys. Rev. Lett.}\ }\textbf {\bibinfo
  {volume} {101}},\ \bibinfo {pages} {221301} (\bibinfo {year}
  {2008})}\BibitemShut {NoStop}%
\end{thebibliography}%

\end{document}